\documentclass[12pt]{article}
\usepackage{amsmath,mathtools}
\usepackage[margin=1in]{geometry}
\usepackage[affil-it]{authblk}
\usepackage{hyperref}
\usepackage{amssymb}
\usepackage{amsthm}
\usepackage{enumitem}
\usepackage{longtable}
\usepackage{amstext}
\usepackage[utf8]{inputenc}
\allowdisplaybreaks[4]
\usepackage{amsthm}

\newtheorem{theorem}{Theorem}[subsection]

\newtheorem{remark}{Remark}[subsection]

\newtheorem*{theorem*}{Theorem}

\usepackage{graphics}

\usepackage{dsfont}
\usepackage{color}
\usepackage{environ}

\allowdisplaybreaks[1]

\newcommand{\R}{\mathbb{R}}
\newcommand{\N}{\mathbb{N}}
\newcommand{\Z}{\mathbb{Z}}

\begin{document}
\title{User manual and tutorial for \texttt{ISIM1s}: \\ a tiny MATLAB package for single stage invariant manifold-guided impulsive stabilization of delay equations}
\author{Kevin E.M. Church
\thanks{Email: kevin.church@mcgill.ca}}
\affil{Department of Mathematics and Statistics\\ McGill University}
\date{}
\clearpage\maketitle
\thispagestyle{empty}

\abstract{\texttt{ISIM1s} consists of a few MATLAB functions and a script that can be used to derive stabilizing impulsive controllers for delay differential equations. This document serves as both a manual and tutorial on the functionality of the \texttt{ISIM1s} package. Brief background on the theoretically guaranteed stabilization scenario are provided before the primary MATLAB script is explained. The tutorial demonstrates how the package can be used to derive stabilizing impulsive controllers for delay differential equations of various complexity scales. Emphasis is placed on the role of various tuning parameters.}

\subsubsection*{Citation.}
Scientific publications for which the package \texttt{ISIM1s} has been used shall
mention its usage and cite the following
publication(s) to ensure proper attribution and reproducibility:\\
\\
\emph{This manual:}\\
Kevin E.M. Church, \textit{User manual and tutorial on \texttt{ISIM1s}: a tiny MATLAB package for single stage invariant manifold-guided impulsive stabilization of delay equations. \emph{https://arxiv.org/abs/1912.07766}}\\
\\
\emph{Theoretical background concerning the trivial unstable subspace case:}\\
Kevin E.M. Church and Xinzhi Liu, \textit{Cost-effective robust stabilization and bifurcation suppression}, SIAM Journal on Control and Optimization, 57 (3), 2240-2268 (2019).\\

\subsubsection*{License.}
\footnotesize BSD 2-Clause license --- \\
\\
Copyright \copyright  2019, Kevin Church\\
All rights reserved.\\
\\
Redistribution and use in source and binary forms, with or without
modification, are permitted provided that the following conditions are met:

\begin{enumerate}
\item Redistributions of source code must retain the above copyright notice, this
   list of conditions and the following disclaimer.
\item Redistributions in binary form must reproduce the above copyright notice,
   this list of conditions and the following disclaimer in the documentation
   and/or other materials provided with the distribution.
\end{enumerate}

\noindent THIS SOFTWARE IS PROVIDED BY THE COPYRIGHT HOLDERS AND CONTRIBUTORS ``AS IS" AND
ANY EXPRESS OR IMPLIED WARRANTIES, INCLUDING, BUT NOT LIMITED TO, THE IMPLIED
WARRANTIES OF MERCHANTABILITY AND FITNESS FOR A PARTICULAR PURPOSE ARE
DISCLAIMED. IN NO EVENT SHALL THE COPYRIGHT OWNER OR CONTRIBUTORS BE LIABLE FOR
ANY DIRECT, INDIRECT, INCIDENTAL, SPECIAL, EXEMPLARY, OR CONSEQUENTIAL DAMAGES
(INCLUDING, BUT NOT LIMITED TO, PROCUREMENT OF SUBSTITUTE GOODS OR SERVICES;
LOSS OF USE, DATA, OR PROFITS; OR BUSINESS INTERRUPTION) HOWEVER CAUSED AND
ON ANY THEORY OF LIABILITY, WHETHER IN CONTRACT, STRICT LIABILITY, OR TORT
(INCLUDING NEGLIGENCE OR OTHERWISE) ARISING IN ANY WAY OUT OF THE USE OF THIS
SOFTWARE, EVEN IF ADVISED OF THE POSSIBILITY OF SUCH DAMAGE.
\normalsize

\subsubsection*{Installation.}
Download \texttt{ISIM1s} from \emph{https://www.kevinchurch.ca/matlab-code.html}. Extract the zip file to a folder of your choice. Download the function \texttt{cheb.m} from \emph{https://people.maths.ox.ac.uk/trefethen/spectral.html} and add it to the folder. Add this folder and subfolders to your MATLAB path.

\section{Introduction}
\texttt{ISIM1s} (Impulsive Stabilization by Invariant Manifold - 1 stage) is a small MATLAB package containing an implementation of the invariant manifold-guided impulsive stabilization method for delay differential equations first proposed by Church and Liu \cite{Church2019b} for systems at bifurcation points and recently extended to the general unstable case \cite{Church_IEETAC}. For detailed theoretical background, we refer the reader to the previous two publications. A sketch of the method follows.

\subsection{The setup}
The starting point is a potentially nonlinear $n$-dimensional delay differential equation
\begin{align*}
\dot y&=f(y(t),y(t-\tau))
\end{align*}
with a single delay $\tau>0$ and an equilibrium $y^*\in\R^n$. It is assumed that $f$ is at least twice continuously differentiable. The goal is to find a function $g$ such that $y^*$ remains an equilibrium in the impulsive system
\begin{align}
\label{NL1}\dot y&=f(y(t),y(t-\tau)),&t&\notin\frac{1}{h}\Z\\
\label{NL2}\Delta y&=g(y(t^-)),&t&\in\frac{1}{h}\Z
\end{align}
and is locally asymptotically stable. We assume that $g$ does not depend on the delayed state\footnote{A future version of \texttt{ISIM1s} will allow for $g$ to depend on the delayed state, at which point this manual will be updated.}, but this assumption is made primarily for ease of presentation. Here, $h>0$ is the frequency of impulse effect. From the theory of impulsive functional differential equations \cite{Church2019a}, the equilibrium is locally asymptotically stable if and only if the same is true of the \emph{linearization}
\begin{align*}
\dot z&=D_1f(y^*,y^*)z(t) + D_2f(y^*,y^*)z(t-\tau),&t&\notin\frac{1}{h}\Z\\
\Delta z&=Dg(y^*)z(t^-),&t&\in\frac{1}{h}\Z,
\end{align*}
where $D_1$ and $D_2$ denote the partial Fr\'echet derivative operators (i.e.\ the operator that maps to the Jacobian matrix) in the first and second variable, and $D$ is the Fr\'echet derivative. If the goal is to \emph{construct} $g$ such that local asymptotic stability is ensured, we see that it is only necessary to construct the matrix $B:=Dg(y^*)$, taking as data the input matrices $A_0:=D_1f(y^*,y^*)$ and $A_1:=D_2f(y^*,y^*)$, the delay $\tau>0$ and the frequency $h>0$ of impulse effect. That is, one \emph{designs} $B\in\R^{n\times n}$ such that
\begin{align}
\label{LIN1}\dot x&=A_0x(t)+A_1x(t-\tau),&t&\notin\frac{1}{h}\Z\\
\label{LIN2}\Delta x&=Bx(t^-),&t&\in\frac{1}{h}\Z
\end{align}
is locally asymptotically stable. Once this is accomplished, the linear-order controller
\begin{align}
\label{gfunction}g(y) = B(y-y^*) + O(|y-y^*|^2)
\end{align}
guarantees local asymptotic stability of \eqref{NL1}--\eqref{NL2}.

\subsection{Invariant manifold-guided impulsive stabilization}
The eigenvalues of the linear system without impulses
\begin{align}\label{LIN1-0}\dot x=A_0x(t)+A_1x(t-\tau)\end{align}
are a subset $\Sigma$ of the complex plane such that $\lambda\in\Sigma$ if and only if $e^{t\lambda}z$ is solution of \eqref{LIN1-0} for some nonzero $z\in\mathbb{C}^n$. This set is generally infinite, but there are only finitely many elements to the right of any vertical line. Eigenvalues with positive real part contribute to instability of \eqref{LIN1-0}, while those with zero real part are responsible for small bounded oscillations or solutions exhibiting sub-exponential growth.

The invariant manifold-guided stabilization procedure is as follows. First, we take the functional form \eqref{gfunction} for the nonlinearity $g$ in \eqref{NL1}--\eqref{NL2} and interpret $B\in\R^{n\times n}$ as a small perturbation parameter. Let $\Sigma_{cu}\subset\Sigma$ denote the eigenvalues of \eqref{LIN1-0} with non-negative real part. Then, the system with impulses \eqref{NL1}--\eqref{NL2} possesses a parameter-dependent centre-unstable manifold at $(y,B)=(y^*,0)$. The dimension of this manifold is $d+n^2$, where $d$ is the dimension of the centre-unstable eigenspace of \eqref{LIN1-0} and the $n^2$ comes from the parameter $B\in\R^{n\times n}$. For details on invariant manifold theory for impulsive function differential equations, one may consult the author's doctoral thesis \cite{Church2019a} or the monograph \cite{ChurchLiu_monograph}. The dynamics on the nontrivial part of the centre-unstable manifold (i.e.\ on the $d$-dimensional slice of the manifold for fixed, small parameter $B$) are smoothly equivalent to those of
\begin{align}
\label{LIN-man1}\dot u&=\Lambda u,&t&\notin\frac{1}{h}\Z\\
\label{LIN-man2}\Delta u&=\Gamma\Psi(0)B\Phi(0)u(t^-),&t&\in\frac{1}{h}\Z,
\end{align}
where $\Phi$ is a real array whose columns form a basis for the ($d$-dimensional) centre-unstable eigenspace, $\Psi$ is an array whose rows form a basis for the tranpose (formal adjoint) centre-unstable eigenspace, $\Lambda\in\R^{d\times d}$ satisfies $\frac{d}{d\theta}\Phi(\theta)=\Phi(\theta)\Lambda$, and $\Gamma=\langle \Psi,\Phi\rangle_\mathcal{C}^{-1}$ is a normalization factor, where 
$$\langle \Psi,\Phi\rangle_\mathcal{C} = \Psi(0)\Phi(0) - \int_0^\tau \Psi(s)A_1\Phi(s-\tau)ds,$$
\vspace{-0.75cm}
\begin{align*}
\mathcal{C}&=\{\phi:[-\tau,0]\rightarrow\R^n : \phi\mbox{ is continuous}\},\\
\mathcal{C}^*&=\{\psi:[0,\tau]\rightarrow\R^{n*} : \psi\mbox{ is continuous}\}.
\end{align*}
Let $\gamma'$ denote the spectral radius of the \emph{monodromy matrix}
\begin{align}\label{monodromy}\mathcal{M}=\left(I+\Gamma\Psi(0)B\Phi(0)\right)e^{\frac{1}{h}\Lambda}.\end{align}
The main theoretical result is as follows.
\begin{theorem}
There exists some $\epsilon>0$ such that if $||B||<\epsilon$, every solution $x(t)$ of the linear system \eqref{LIN1}--\eqref{LIN2} satisfies the exponential estimate $||x(t)||\leq C||x_0||e^{r t}$ for some constant $C$, with $r=h\log(\gamma') + O(||B||^2)$. 
\end{theorem}

Invariant manifold-guided impulsive stabilization is therefore not so much determined by the invariant manifold itself, but by the topologically equivalent linear-order dynamics \eqref{LIN-man1}--\eqref{LIN-man2} on this manifold. The heuristic one may therefore follow to generate a stabilizing controller (matrix) $B$ is as follows.
\begin{itemize}
\item Choose a frequency parameter $h$ and a target convergence rate parameter $\gamma$.
\item Compute the relevant matrices $\Gamma$, $\Phi(0)$ and $\Psi(0)$ needed in the computation of the monodromy matrix $\mathcal{M}$.
\item Find a (small) matrix $B^*$ such that $\rho(\mathcal{M})\leq e^{\frac{1}{h}\gamma}$
\item The local convergence rate of \eqref{NL1}--\eqref{NL2} with $g(y)=B^*(y-y^*)+O(|y-y^*|^2)$ is $O(e^{rt})$ with $r=h\log\left(e^{\frac{1}{h}\gamma}\right)=\gamma + O(||B||^2)$, provided all eigenvalues with negative real part of the delay differential equation have real part strictly less than $\gamma$.
\end{itemize}
Note that $\rho(X)$ is the spectral radius of the matrix $X$. This heuristic is precisely the basis for the package \texttt{ISIM1s}. The reference to ``one stage" in the name of the package (the \texttt{1s} part of \texttt{ISIM1s}) refers to the fact that the procedure only involves one linearization step. The theoretical justification for this heuristic appears in the papers \cite{Church2019b} and \cite{Church_IEETAC}.

One can impose constraints on the form of the matrices $B$. For example, one might require that these must lie in a particular convex subset of the $n\times n$ matrices. One can also include a performance target, such as requiring $B^*$ minimize a suitable cost functional. This is incorporated into \texttt{ISIM1s} and the theoretical feasibility of such more general problems are covered in \cite{Church2019b}.

It may be that your system contains some centre-stable modes that are inaccessible or can not be controlled. As such, you might want to control those eigenvalues with strictly positive real part while ignoring those with zero real part. Alternatively, you might have some periodic solutions that can not be controlled, but you want to improve the convergence rate toward such solutions by altering the transient dynamics in a suitable way. By controlling those eigenvalues with negative real part and requesting a better convergence rate from \texttt{ISIM1s}, this can be accomplished.

\subsection{Ways the procedure can fail}\label{section-fail}
We prefaced the above procedure by describing it as a heuristic. This is because there are several reasons it may fail. Chief among these is that the matrix $B^*$ that ensures the spectral radius requirement $\sigma(\mathcal{M})\leq e^{\frac{1}{h}\gamma}$ might be too large. At a theoretical level, the heuristic breaks down because of two related problems.
\begin{enumerate}
\item The linear-order dynamics on the parameter-dependent centre-unstable manifold are no longer described by \eqref{LIN-man1}--\eqref{LIN-man2} because $||B^*||$ is too large.
\item Some eigenvalue\footnote{More precisely, Floquet exponent} of \eqref{LIN1}--\eqref{LIN2} with negative real part crosses the imaginary axis along some (matrix) parameter curve $\mu\mapsto B(\mu)$ with $B(0)=0$ and $B(1)=B^*$.
\end{enumerate}
The first one of these is technical; the centre-unstable manifold is defined by way of a cutoff procedure, so the dynamics ``far away" from $(y,B)=(y^*,0)$ on this manifold might not be reflective of the local representation provided by the dynamics equation \eqref{LIN-man1}--\eqref{LIN-man2}. Whether this technical problem really has an impact is unclear. The quadratic error $O(||B||^2)$ may also start to dominate the controlled part of the spectrum when the matrix $B$ becomes large.

The second of these two points of failure, however, is a fair bit easier to understand. Since the above procedure is based on only controlling the unstable (and centre) modes of the delay differential equation, it is conceivable that attempting to stabilize these modes by way of impulses might result in a stable mode becoming unstable. This is apparent even in the finite-dimensional setting, as we will see in the tutorial. The good news is that this point of failure can sometimes be mitigated by \emph{also} controlling stable modes with near-zero real parts. In effect, one extends the procedure to the (parameter-dependent) centre-unstable manifold taken in union with a finite-dimensional portion of the stable manifold corresponding to those troublesome stable modes. This can be accounted for in \texttt{ISIM1s}. 

\section{Overview of \texttt{ISIM1s}}

\subsection{Dependencies}
\texttt{ISIM1s} makes use of the smooth constrained optimization solver \texttt{fmincon} from the MATLAB Optimization Toolbox. The Global Optimization Toolbox is needed to use the black box  \texttt{patternsearch} solver and the genetic algorthm \texttt{ga}. 

\subsection{Overview of individual functions involved in stabilization}
The following is an overview of the individual MATLAB functions, in order of appearance in which they are called by the main script \texttt{ISIM1s} (or in which their execution is fully completed by said script).
\subsubsection{\texttt{cheb.m}}
Generates the Chebyshev differentiation matrix used in the discretization of the infinitesimal generator of the linear DDE \eqref{LIN1-0}. This code appears in the book \cite{Trefethen2000} and is also publicly available on the author's website. This \texttt{m} file is not included in \texttt{ISIM1s.zip} and must be downloaded by the user from the website listed in the installation section at the beginning of this document.

\subsubsection{\texttt{dde\_data.m}}
 This function computes the matrices $\Lambda$, $\Gamma$, $\Phi(0)$ and $\Psi(0)$. It is assumed that for each eigenvalue $\xi$, the
 dimension of the kernel of the characteristic matrix 
 $$\Delta(\xi)=I\xi - A_0 - A_1e^{-\xi\tau}$$ is equal to its multiplicity. 
 If this condition is not satisfied, all outputs will be incorrect. This function includes a call to \texttt{cheb.m}. The eigenvalues are computed by discretizing the infinitesimal generator associated to the delay differential equation. The implementation is quite efficient, taking only a few lines of code \cite{Jarlebring2008}.
 
The columns of $\Phi(0)$ (and rows of $\Psi(0)$) are obtained by computing the eigenvalues and eigenvectors of the discretized infinitesimal generator of the DDE. It is assumed that the generalized eigenspaces associated to each eigenvalue are spanned by rank 1 eigenvectors.\footnote{A future version of \texttt{ISIM1s} will include the option to instead use the \texttt{jordan} decomposition (at the cost of speed, since \texttt{jordan} operates at a symbolic level as opposed to with floating point arithmetic). That update will provide a means of removing the assumption on the generalized eigenspaces. At that point, this manual will be updated.} If this assumption does not hold, the outputs might be incorrect.\\
\\
\emph{Inputs:}
\begin{itemize}
\item \texttt{A0}: the matrix $A_0$ in \eqref{LIN1-0}.
\item \texttt{A1}: the matrix $A_1$...
\item \texttt{tau}: the delay $\tau$...
\item \texttt{N}: number of Chebyshev nodes in the differentiation matrix. Requires $\texttt{N}\geq\texttt{2}$.
\item \texttt{eig\_lower}: eigenvalues of \eqref{LIN1-0} with real part less than \texttt{eig\_lower} will be ignored.
\item \texttt{eig\_upper}: eigenvalues of \eqref{LIN1-0} with real part greater than \texttt{eig\_upper} will be ignored. Set $\texttt{eig\_upper}=\texttt{inf}$ if you want to include all eigenvalues with real part greater than \texttt{eig\_lower}.
\end{itemize}

\noindent\emph{Outputs:}
\begin{itemize}
\item \texttt{LAM}: the matrix $\Lambda$
\item \texttt{Phi0}: the matrix $\Phi(0)$
\item \texttt{Psi0}: the matrix $\Psi(0)$
\item \texttt{GAM}: the matrix $\Gamma$
\item \texttt{eigs\_all}: the complete list of eigenvalues. Useful for parameter tuning purposes to check if some stable eigenvalue is at risk of crossing into the right half-plane.
\end{itemize}

\begin{remark}
If \emph{\texttt{eig\_lower}} and \emph{\texttt{eig\_upper}} are not specified or are both empty -- that is, only the first four inputs are passed to \emph{\texttt{dde\_data}} or the final two of six are input as \emph{\texttt{{[]}}} -- then the outputs \emph{\texttt{LAM,Phi0,Psi0,GAM}} will not be computed. This is useful if you are working with a very high-dimensional system and want to determine the approximate location (and real parts) of your eigenvalues by using a small number \emph{\texttt{N}} of discretization nodes before prescribing the search range \emph{\texttt{{[}eig\_lower,eig\_upper{]}}} and computing the rest of the data with a higher number of nodes, which might be very expensive.
\end{remark}

\subsubsection{\texttt{probematrix.m}}
Relative to the decomposition $\mathcal{M}=M_0(B)+Z$ of the monodromy matrix \eqref{monodromy} with $$Z=e^{\frac{1}{h}\Lambda},\hspace{1cm}M_0(B)=\Gamma\Psi(0)B\Phi(0)Z,$$
this program computes the matrix $Z$ and the matrix of $B\mapsto M_0(B)$ relative to the input \texttt{basis}.\\
\\
\emph{Inputs:}
\begin{itemize}
\item \texttt{LAM,Phi0,Psi0,GAM}: outputs from \texttt{dde\_data.m}
\item \texttt{h}: positive real frequency parameter $h$ for impulsive stabilization
\item \texttt{basis}: a $n\times(n\cdot k)$ matrix such that $\texttt{basis}_i:=\texttt{basis(:,(i-1)*n+1:i*n)}$ is $i$th basis element (matrix) for the particular $k$-dimensional subspace of $n\times n$ matrices you want to use in control synthesis
\end{itemize}

\noindent\emph{Outputs:}
\begin{itemize}
\item \texttt{M0}: a $d\times(d\times k)$ matrix such that $\texttt{M0(:,(i-1)*d+1:i*d)}=M_0(\texttt{basis}_i)$.
\item \texttt{Z}: the numerical matrix exponential, $\exp(\frac{1}{h}\Lambda)$.
\item \texttt{M0\_vectorized,Z\_vectorized}: reshaped $d^2\times k$ and $d^2\times 1$ versions of the matrices \texttt{M0} and \texttt{Z}.
\end{itemize}

\begin{remark}
One could instead encode basis information into a constraint function (see \emph{\texttt{optimize.m}}) and use the standard basis for $\R^{n\times n}$ here, but for a very high-dimensional problem this could slow down the optimization step. We have therefore allowed a basis to be specified explicitly.
\end{remark}

\subsubsection{\texttt{optimize.m}}
Solves the optimization problem
\begin{align*}
\begin{aligned}
\mbox{minimize }&\quad\mathcal{C}_1(B),\\
\mbox{subject to }&\quad \rho(\mathcal{M}(B))\leq e^{\gamma/h},\hspace{1mm} c(B)\leq 0,
\end{aligned}
\end{align*}
where $c:\mathcal{U}\rightarrow\R$ is convex and $\mathcal{U}=\mbox{span}\{\texttt{basis}_1,\dots,\texttt{basis}_m\}.$ The problem with $c\equiv 0$ is provably feasible provided the rank of the linear function $M_0:\mathcal{U}\rightarrow\R^{d\times d}$ is at least $d^2$. In this case, the optimization will be completed in the probe space (i.e.\ the image of $\mathcal{M}:\mathcal{U}\rightarrow\R^{d\times d}$) provided the dimension of the range ($d^2$) is smaller than that of the control space ($k$) and the cost function $\mathcal{C}_1$ is quadratic. This is done for efficiency, and the explicit transformation that preserves the cost function is explained in \cite{Church2019b}.\\
\\
\emph{Inputs:}
\begin{itemize}
\item \texttt{M0\_vectorized,Z\_vectorized,h,basis}: Outputs and inputs (same names) from previous programs
\item \texttt{constraint}: an anonymous function $c:\R^k\rightarrow\R\ell$ describing the constraint on $\mathcal{U}=\mbox{span}\{\texttt{basis}_i:i=1,\dots,k\}$ in terms of coordinates in $\R^k$. If there is no constraint, set $\texttt{constraint}=\texttt{[]}$.
\item \texttt{cost\_weight,cost\_general}: for a cost function of the form $$\mathcal{C}_1(x) = x^\intercal Wx + C(x)$$ for $x\in\R^k$ representing coordinates of the basis $\mathcal{U}$ with $W\in\R^{k\times k}$ symmetric and positive-definite and $C$ a general positive definite function, choose $\texttt{cost1\_weight}=W$ and input \texttt{cost1\_general} as the anonymous function $C(x)$. If $C(x)\equiv 0$, this should be input as $\texttt{cost1\_general}=\texttt{[]}$. Note also that the unweighted quadratic cost $x\mapsto x^\intercal x$ can be input using \texttt{cost\_weight = 1}, \texttt{cost\_general = []}.
\item \texttt{gamma}: real number $\gamma$ for the target local convergence rate $O(e^{\gamma t})$.
\item \texttt{searchmode}: specifies the built-in MATLAB solver for the optimization step. \texttt{`fmincon'} is the local smooth\footnote{Warning, the input problem is not smooth because of the spectral radius constraint (and if any of your cost functions or constraints are nonsmooth). If using `fmincon', any numerical local minimum that is found will only be guaranteed to be such (to standard tolerances) if the eigenvalues of the monodromy matrix are simple at this point. To be safe, consider using the local minimum as a guess for \texttt{`patternsearch'} mode.} solver, \texttt{`patternsearch'} is the deterministic black box pattern search solver, and \texttt{`ga'} is the nondeterminstic genetic algorithm \texttt{ga}.
\item \texttt{guess}: An initial guess for the solver. This is only relevant for \texttt{`fmincon'} and \texttt{`patternsearch'} search modes. If $\texttt{guess}=\texttt{[]}$, the guess will be taken as a random vector.
\item \texttt{options}: An options structure for the relevant solver. See relevant MATLAB documentation. Set to empty \texttt{[]} if you want to use default options.
\item Note: if no \texttt{searchmode}, \texttt{guess} or \texttt{options} data is specified (i.e.\ the function is only given eight inputs) the solver will default to \texttt{`ga'} with empty options structure.
\end{itemize} 

\noindent\emph{Outputs:}
\begin{itemize}
\item \texttt{x}: coordinate in $\R^k$ of the numerical solution of the optimization problem
\item \texttt{fval}: cost of the control.
\end{itemize}

\begin{remark}
The cost functions, weight matrix and constraint function are all input relative to the \emph{coordinates} for control elements in terms of the specified $\texttt{\emph{basis}}$. 
\end{remark}

\subsection{The main script: \texttt{ISIM1s\_run.m}}
The script \texttt{ISIM1s\_run.m} runs the previous programs in order given numerous user inputs. The end result is a pair \texttt{[control,cost]} with the first element being a cell array such that \texttt{cost\{1\}} is the coordinate in the basis $\mathcal{U}$ of the identified optimal control, \texttt{control\{2\}} is the control as an $n\times n$ matrix, and \texttt{cost} is its associated cost. The user inputs are as follows.
\begin{itemize}
\item \texttt{A0,A1,tau}: The delay differential equation data.
\item \texttt{control\_basis}: A basis for subspace of control matrices on which any constraints will be specified; input in the same way as an input to \texttt{probematrix.m}.
\item \texttt{h,gamma}: Frequency of impulse effect (\texttt{h}) and target convergence/growth rate parameter.
\item \texttt{constraint,cost\_weight,cost\_general,searchmode,options}: See inputs to \texttt{optimize.m}.
\item \texttt{eig\_lower,eig\_upper}: Same as the parameters from \texttt{dde\_data.m}; specifies a cutoff window for the eigenvalues to control for. 
\item \texttt{spectrum\_discretization}: Same as the parameter $\texttt{N}\geq\texttt{2}$ from \texttt{dde\_data.m}; specifies number of Chebyshev modes.
\item \texttt{refining}: If set to \texttt{`true'}, \texttt{dde\_data.m} and \texttt{probematrix.m} will not run. To be used (to save computation time) if you previously ran \texttt{ISIM1s\_run.m} and your input delay DE data (\texttt{A0,A1,tau}), control input data (\texttt{control\_basis,h}) and spectrum parameters (\texttt{eig\_lower,eig\_upper,spectrum\_discretization}) are not going to be changed in subsequent runs. Useful for working with sequences of objective functions for discontinuous cost functions or to validate a \texttt{`fmincon'} search mode solution by taking it as a guess for \texttt{`patternsearch'} mode.
\end{itemize}

\subsection{Convenience function for plotting/testing}
Included is a function to facilitate the verification of whether a controller identified by \texttt{ISIM1s\_run} results in stabilization.
\subsubsection{\texttt{testing\_matrixmode.m}}
\emph{Inputs:}
\begin{itemize}
\item \texttt{A0,A1,tau,h}: same interpretation as in previous function files.
\item \texttt{control}: a matrix with the same dimensions as \texttt{A0}. This should typically be the output \texttt{control\{2\}} from \texttt{ISIM1s\_run} if you want to test the identified controller for stabilization.
\item \texttt{pulselimit}: rather than specifying the length of time for the simulation, here you specify the number of impulses. If the frequency is $h$, then the simulation time will be $[0,\texttt{pulselimit/h}]$.
\end{itemize}
\emph{Outputs:}
\begin{itemize}
\item \texttt{sol}: The solution through the constant initial condition $x_0=(1,\dots,1)$ of the system \eqref{LIN1}--\eqref{LIN2} presented as a \texttt{dde23} solution structure. 
\end{itemize}

\section{Tutorial}
This tutorial will demonstrate how the script \texttt{ISIM1s\_run.m} can be used to stabilize a delay differential equation. It should also be helpful in explaining how a basis for the control space should be input. We will see a few instances where the script fails to generate a stabilizing controller; these will be useful learning exercises and in these cases we will show how modifying the \texttt{frequency} and/or \texttt{eig\_lower} parameter can sometimes make the script generate a functioning controller. The first five sections are concrete examples, and some general conditioning guidelines follow in Section \ref{conditioning}.

\emph{Format:} In these sections, modifications to the user data of \texttt{ISIM1\_run.m} will be stated in teletype font. The symbol \texttt{>>} indicates a MATLAB execution, with subsequent lines displaying MATLAB output and additional executions if applicable. The \texttt{`fmincon'} solver will usually be used, but similar results should result if you use the exact same inputs except you use the \texttt{`patternsearch'} solver instead. Genetic algorithm is nondeterministic, so your output might be drastically different than what is shown here if you use \texttt{`ga'}.

\subsection{A finite-dimensional system}
Consider the two-dimensional system
\begin{align}
\dot x&=\left[\begin{array}{cc}1&0\\0&-1 \end{array}\right]x.\label{ex1-1-ode}
\end{align}
The unstable subspace is spanned by $\texttt{[1;0]}$, the transpose unstable subspace is spanned by its transpose, $\Lambda=1$ and $\Gamma=1$. It should be clear that to stabilize this ordinary differential equation, we need to control the $x_1$ component. Suppose we work with the (impulsive) control space \begin{align}\label{ex1-basis1}\mathcal{U}=\mbox{span}\{\texttt{[-1,0;0,2],[0,0;0,1]}\}.\end{align}
then the dynamics on the parameter-dependent unstable manifold are topologically equivalent to
\begin{align}
\label{CM-dyn-ex1}\dot z&=z,&t&\notin\frac{1}{h}\Z\\
\label{CM-dyn-ex2}\Delta z&=B_{11}z(t^-)&t&\in\frac{1}{h}\Z,
\end{align}
for $B\in\mathcal{U}$. The second basis vector has no effect on the dynamics, but the first one does. As such, in an attempt to minimize the given cost functional, the optimal controller will be one of the form $\texttt{q[-1,0;0,2]}$ for some $\texttt{q}>0$. However, this controller could also destabilize the second component.

\begin{figure}
\centering
\includegraphics[scale=0.7]{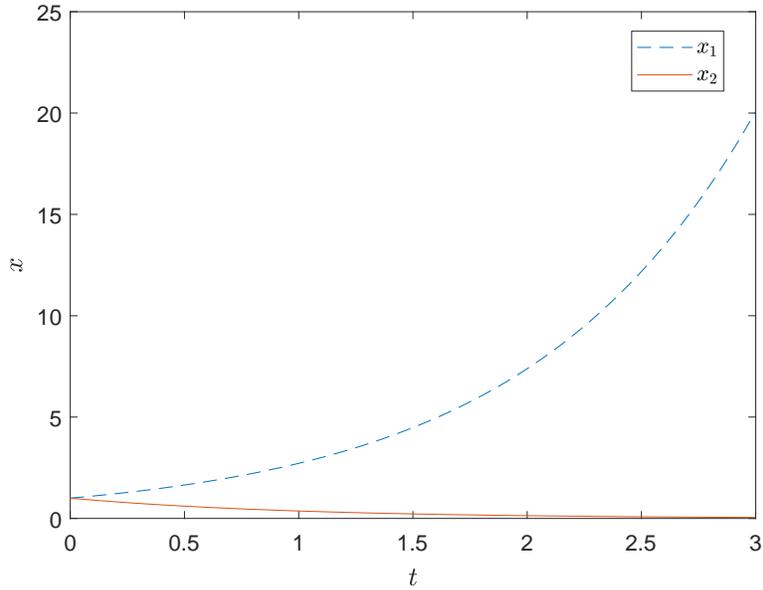}
\caption{Solution of \eqref{ex1-1-ode} without any impulsive control from $x(0)=(1,1)$.}\label{fig-ex1-1}
\end{figure}

Let us explore this problem with \texttt{ISIM1s}. The solution from the initial condition $x(0)=(1,1)$ without any control is plotted in Figure \ref{fig-ex1-1}. One can see that $x_2$ decays to zero and $x_1$ grows exponentially. 
Input the following user data in \texttt{ISIM1s\_run.m}:\\
\\
\noindent\texttt{A0=[1,0;0,-1];\\
A1=zeros(2,2);\\
tau=1; \% set as arbitrary but nonzero\\
control\_basis = [[-1,0;0,2],[0,0;0,1]];\\
h = 1;\\
gamma = -0.2;\\
constraint = [];\\
cost\_weight = diag([1,1]);\\
cost\_general = [];\\
searchmode = `fmincon';\\
guess = zeros(2,1); \\
options = [];\\
eig\_lower = 0;\\
eig\_upper = inf;\\
spectrum\_discretization = 10;\\
refining = `false';}\\
\\
\indent We are still prioritizing a ``small" controller with this choice \texttt{cost\_weight}, but have left the problem unconstrained. The cost function is equivalent to the unweighted quadratic cost $x\mapsto x^\intercal x$. Since \texttt{eig\_lower} is set to zero, the algorithm will not attempt to control for the stable eigenvalue resulting from the $x_2$ component. We run the script.\\
\\
\noindent\texttt{>>ISIM1s\_run\\
Found 1 eigenvalues with real part at least -0.000000 for the\\
input DDE, counting multiplicity.\\
\\
Dimension of control space (2) is at least probe space dimension (1). \\
Unconstrained problem is provably feasible. Performing optimization\\
in probe space.\\
\\
Local minimum found that satisfies the constraints.\\
\\
>>control\{2\}\\
ans = [-0.4512,0;0,0.9024]}\\

\begin{figure}
\centering
\includegraphics[scale=0.7]{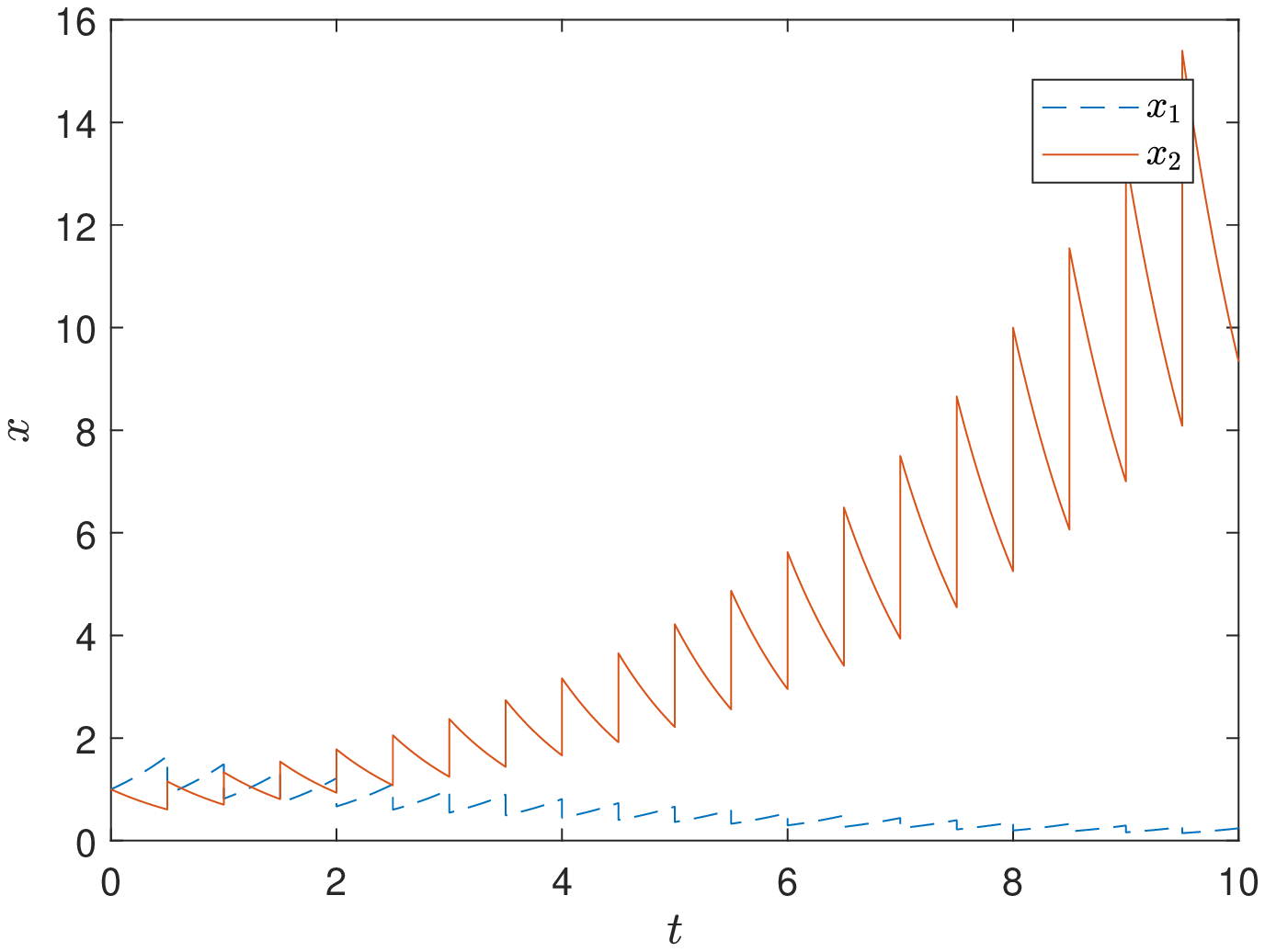}
\caption{Solution of \eqref{ex1-2-ode}--\eqref{ex1-2-jump} with $\texttt{control\{2\}}=\texttt{[-0.4512,0;0,0.9024]}$ and $\texttt{h}=1$ from $x(0)=(1,1)$.}\label{fig-ex1-2}
\end{figure}

The controller is precisely $0.4512\cdot\texttt{[-1,0;0,2]}$, with the latter being our first basis vector. As we know from previous work, this might not yield stability since this controller causes some instability in the second component $x_2$. If we simulate the impulsive differential equation
\begin{align}
\label{ex1-2-ode}\dot x&=\left[\begin{array}{cc}1&0\\0&-1 \end{array}\right]x(t),&t&\notin\frac{1}{h}\Z\\
\label{ex1-2-jump}\Delta x&=\texttt{control\{2\}}x(t^-),&t&\in\frac{1}{h}\Z,
\end{align}
the $x_1$ component is stable but the $x_2$ component is unstable; see Figure \ref{fig-ex1-2}. This is to be expected.

To fix this problem, we will modify the \texttt{eig\_lower} parameter. We know that there is a stable eigenvalue with real part $-1$, so we take \texttt{eig\_lower = -2}. The rest of the user inputs are the same as before. Running the script again, 

\noindent\texttt{>> ISIM1s\_run\\
Found 2 eigenvalues with real part at least -2.000000 for the\\
input DDE, counting multiplicity.\\
\\
Dimension of control space (2) is at most probe space dimension (4).\\
\\
Performing optimization in control space.\\
\\
Local minimum possible. Constraints satisfied.\\
\\
>> control\{2\}\\
ans = [-0.7128,0;0,1.1219]}\\

\begin{figure}
\centering
\includegraphics[scale=0.7]{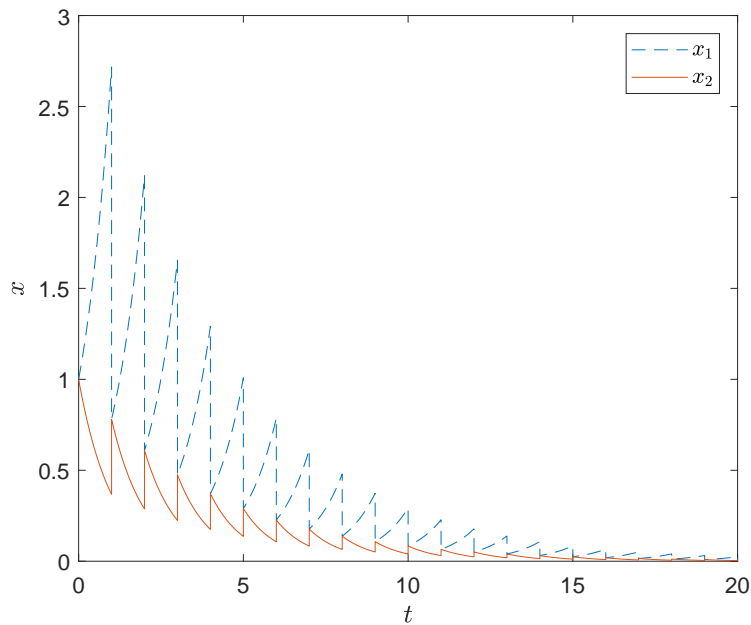}
\caption{Solution of \eqref{ex1-2-ode}--\eqref{ex1-2-jump} with $\texttt{control\{2\}}=\texttt{[-0.7128,0;0,1.1219]}$ and $\texttt{h}=1$ from $x(0)=(1,1)$. Exponential stability is achieved.}\label{fig-ex1-3}
\end{figure}

The new controller can be written in the form 
\begin{align}\label{ex1-Bstar}B^*=\left[\begin{array}{cc}-0.7128&0\\0&1.1219 \end{array}\right]=0.7128\cdot\left[\begin{array}{cc}-1&0\\0&2\end{array}\right] +  (-0.3037)\cdot\left[\begin{array}{cc}0&0\\0&1\end{array}\right],\end{align} which is indeed a linear combination of the prescribed basis elements from \eqref{ex1-basis1}. This time stability is achieved; see Figure \ref{fig-ex1-3}. 

The main message here is that for some stabilization problems, it might be necessary to increase the \texttt{eig\_lower} parameter. In this instance it was somewhat obvious from the differential equation and the choice of control basis that ignoring the stable mode was going to cause problems. However, when delays are involved or the system being controlled is high-dimensional, it might not be as clear.

\subsection{A two-dimensional system with delay}
\begin{figure}
\centering
\includegraphics[scale=0.7]{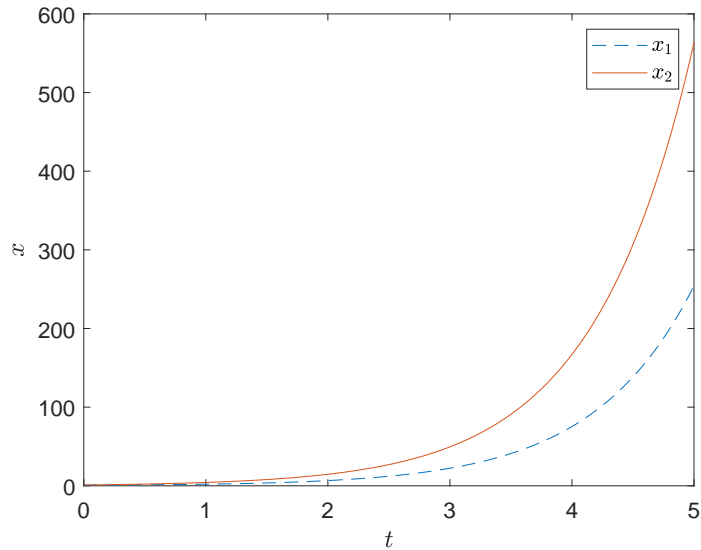}
\caption{Solution of \eqref{ex2-ode}--\eqref{ex2-jump} with $B=0$ from the constant initial condition $x_0=(1,1)$. The trivial solution is unstable.}\label{fig-ex2-1}
\end{figure}

\begin{figure}
\centering
\includegraphics[scale=0.7]{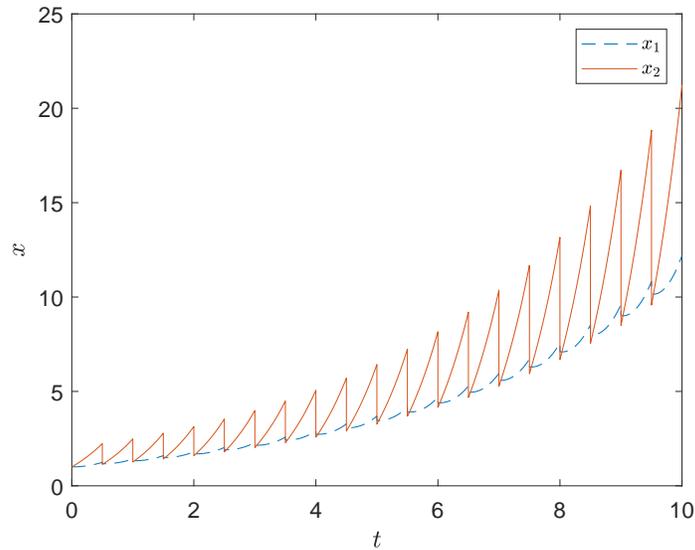}
\caption{Solution of \eqref{ex2-ode}--\eqref{ex2-jump} with $B=\texttt{[-0.0616,0;0,-0.4919]}$ from the constant initial condition $x_0=(1,1)$. We still have instability but the growth rate is not as high.}\label{fig-ex2-2}
\end{figure}

\begin{figure}
\centering
\includegraphics[scale=0.7]{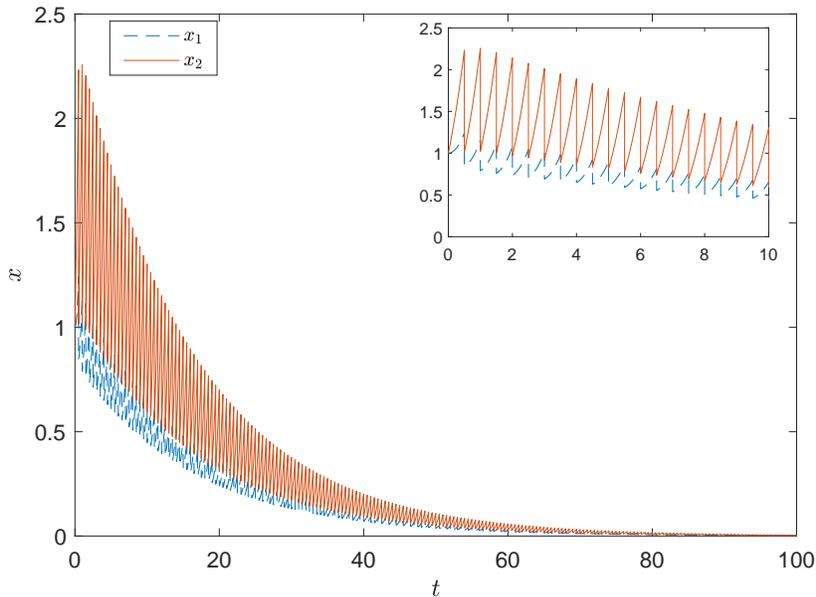}
\caption{Solution of \eqref{ex2-ode}--\eqref{ex2-jump} with $B=\texttt{[-0.3189,0;0,-0.5453]}$ from the constant initial condition $x_0=(1,1)$. Stability is achieved. Inset: windowed solution for smaller time arguments.}\label{fig-ex2-3}
\end{figure}

Consider the stabilization problem 
\begin{align}
\label{ex2-ode}\dot x&=\left[\begin{array}{cc}-1&1\\0&1\end{array}\right]x(t) +
\left[\begin{array}{cc}0&0\\1&-0.1\end{array}\right]x(t-0.4),&t&\notin\frac{1}{2}\Z\\
\label{ex2-jump}\Delta x&=Bx(t^-),&t&\in\frac{1}{2}\Z,
\end{align}
where $B\in\R^{2\times 2}$ is to be designed such that exponential stability is achieved with convergence rate $O(e^{-0.2t})$. We will use the control basis 
\begin{align}\label{ex2-basis1}
\mathcal{U}=\mbox{span}\{[1,0;0,0],[0,0;0,1]\}
\end{align}
for the diagonal subspace. Relative to this basis, we will use the standard quadratic cost functional $C(x)=x^\intercal x$. The inputs to \texttt{ISIM1s\_run.m} are therefore as follows.\\
\\
\noindent\texttt{A0=[-1,1;0,1];\\
A1=[0,0;1,-0.1;\\
tau=0.4; \% set as arbitrary but nonzero\\
control\_basis = [[1,0;0,0],[0,0;0,1]];\\
h = 2;\\
gamma = -0.2;\\
constraint = [];\\
cost\_weight = diag([1,2]);\\
cost\_general = [];\\
searchmode = `fmincon';\\
guess = zeros(2,1); \\
options = [];\\
eig\_lower = 0;\\
eig\_upper = inf;\\\\
spectrum\_discretization = 10;\\
refining = `false';}\\
\\
We have taken \texttt{eig\_lower} to be zero for the time being. Without any control ($B=0$) the system is unstable; see Figure \ref{fig-ex2-1}. 
Running the MATLAB script,\\
\\
\noindent\texttt{>> ISIM1s\_run\\
Found 1 eigenvalues with real part at least -2.000000 for the input DDE, counting multiplicity.\\
\\
Dimension of control space (2) is at least probe space dimension (1).\\
Unconstrained problem is provably feasible. Performing optimization\\
in probe space.\\
\\
Local minimum found that satisfies the constraints.\\
\\
>> control{2}\\
ans = [-0.0616,0;0,-0.4919]\\}
\\
\indent Unfortunately, this controller does not stabilize the system; see Figure \ref{fig-ex2-2}.
Let us examine the eigenvalues of the continuous part, \eqref{ex2-ode}.\\
\\
\noindent\texttt{>> flip(eigs\_all(end-2:end))\\
\\
ans = [1.2160 + 0.0000i;-1.7915 + 0.0000i;-9.8122 + 0.0000i]\\}
\\
As we can see there is one eigenvalue $1.2160$ with positive real part as well as another real eigenvalue $-1.7915$ that is negative. This eigenvalue might have been destabilized by the controller $B$ identified in the previous run of the script. Let us modify the eigenvalue cutoff parameter so that this eigenvalue is accounted for. We set \texttt{eig\_lower = -2} and run the script again.\\
\\
\noindent \texttt{Found 2 eigenvalues with real part at least -2.000000 for the input DDE, counting multiplicity.\\
\\
Dimension of control space (2) is at most probe space dimension (4).\\
\\
Performing optimization in control space.\\
\\
Local minimum found that satisfies the constraints.\\
\\
>> control\{2\}\\
ans = [-0.3189,0;0,-0.5453]\\}
\\
This time around, stability is achieved; see Figure \ref{fig-ex2-3}.
The desired convergence rate of $O(e^{-0.2t})$, however, is not quite attained. With some logarithmic transformation to the solution and basic fitting, we extract a convergence rate that is closer to $O(e^{-0.062t})$. To attempt to overcome this weaker convergence rate, we can further decrease the target convergence rate parameter $\texttt{gamma}$. Choosing $\texttt{gamma}=\texttt{-0.4}$ and running the script again, the result is the controller $\texttt{control\{2\}}=\texttt{[-0.3594,0;0,-0.5911]}$, which can be shown to achieve the convergence rate $O(e^{-0.22t})$.

To summarize, if the script fails to generate a controller that stabilizes your system, it is worth inspecting \texttt{eigs\_all} and adjusting the \texttt{eig\_lower} parameter so that one or more eigenvalues with negative real part is controlled as well. In some cases it might also be helpful to adjust the target convergence rate parameter $\texttt{gamma}$.

\begin{remark}
If your problem requires \texttt{\emph{eig\_lower}} to be quite large, it is also a good idea to increase \texttt{\emph{spectrum\_discretization}} beyond the stock setting of ten. This is because eigenvalues with larger negative real part are more likely to be artifacts of the discretization scheme, and increasing the latter parameter improves accuracy and suppresses these (at the cost of computation time).
\end{remark}

\subsection{A scalar delay differential equation that can not be stabilized at certain frequencies}\label{ex3}
The scalar DDE
\begin{align*}
\dot x&=-\frac{\pi}{2}x(t-1)
\end{align*}
has the eigenvalues $\pm i\frac{\pi}{2}$, with all others (countably many) having negative real part. In particular, the functions $$x_1(t)=\sin\left(\frac{t\pi}{2}\right)\hspace{1cm}x_2(t)=\cos\left(\frac{t\pi}{2}\right)$$ are both solutions. The equilibrium $x=0$ is therefore stable, but not asymptotically stable. 

Consider the impulsive stabilization problem
\begin{align}
\label{ex3-dde1}
\dot x&=-\frac{\pi}{2}x(t-1),&t&\notin\frac{1}{h}\Z\\
\label{ex3-dde2}\Delta x&=bx(t^-),&t&\in\frac{1}{h}\Z,
\end{align}
where we must find $b\in\R$ so that $x=0$ is asymptotically stable. If $h=\frac{1}{2N}$ for some $N\in\N$, this problem has no solutions. To see why this is the case, observe that when $t=k/h=2kN$ for $k\in\Z$ and $h=\frac{1}{2N}$, the two functions $x_1$ and $x_2$ from above satisfy $$x_1(kN^-)=\sin\left(kN\pi\right),\hspace{1cm}x_2(kN^-)=\cos\left(kN\pi\right).$$ Both of these are zero for all $k\in\Z$. Consequently, the impulse effect \eqref{ex3-dde2} does nothing to these solutions. In particular, $x_1(t)$ and $x_2(t)$ are periodic solutions of \eqref{ex3-dde1}--\eqref{ex3-dde2}, so this system is not asymptotically stable. 

\begin{figure}
\centering
\includegraphics[scale=0.45]{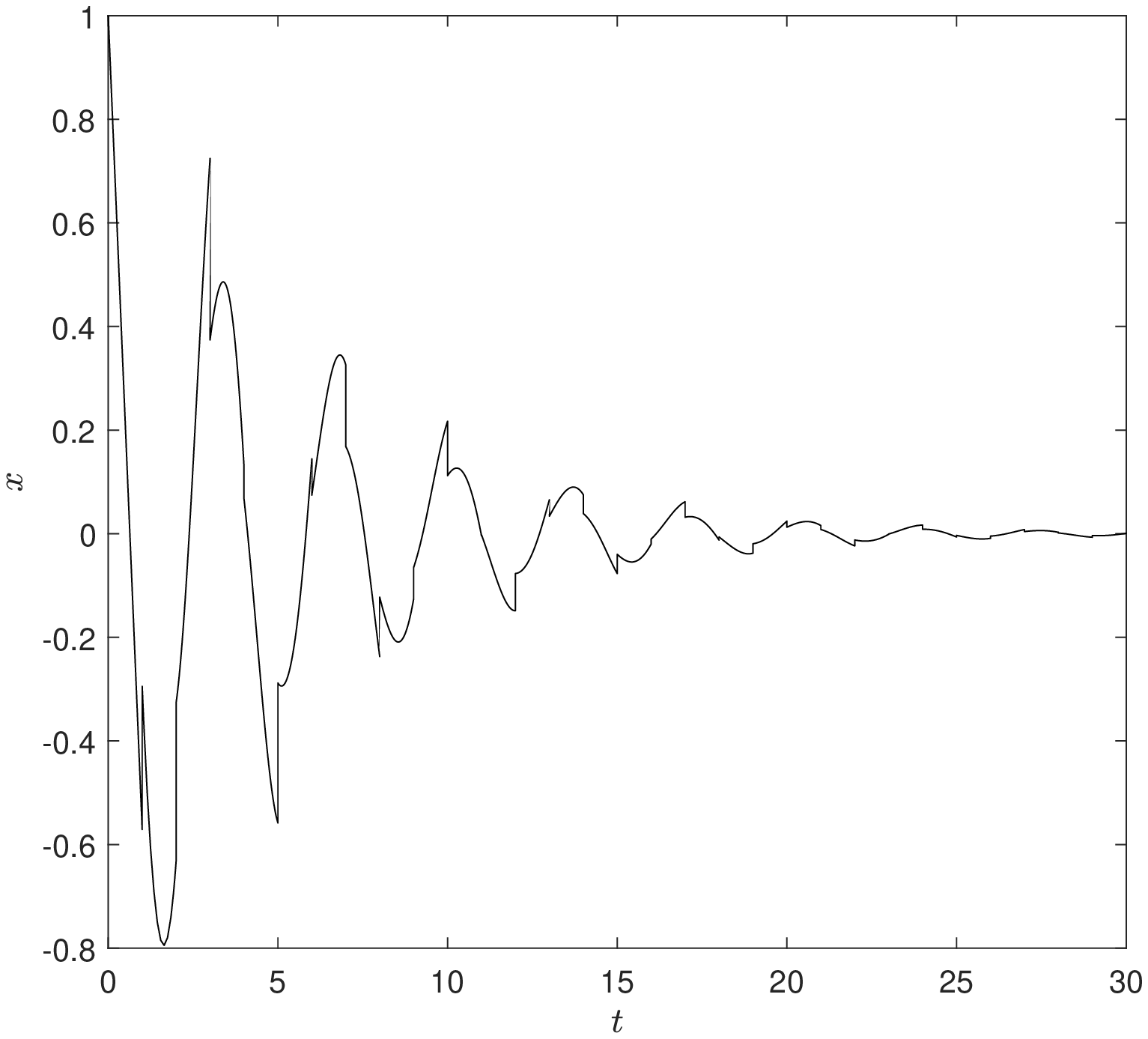}\includegraphics[scale=0.45]{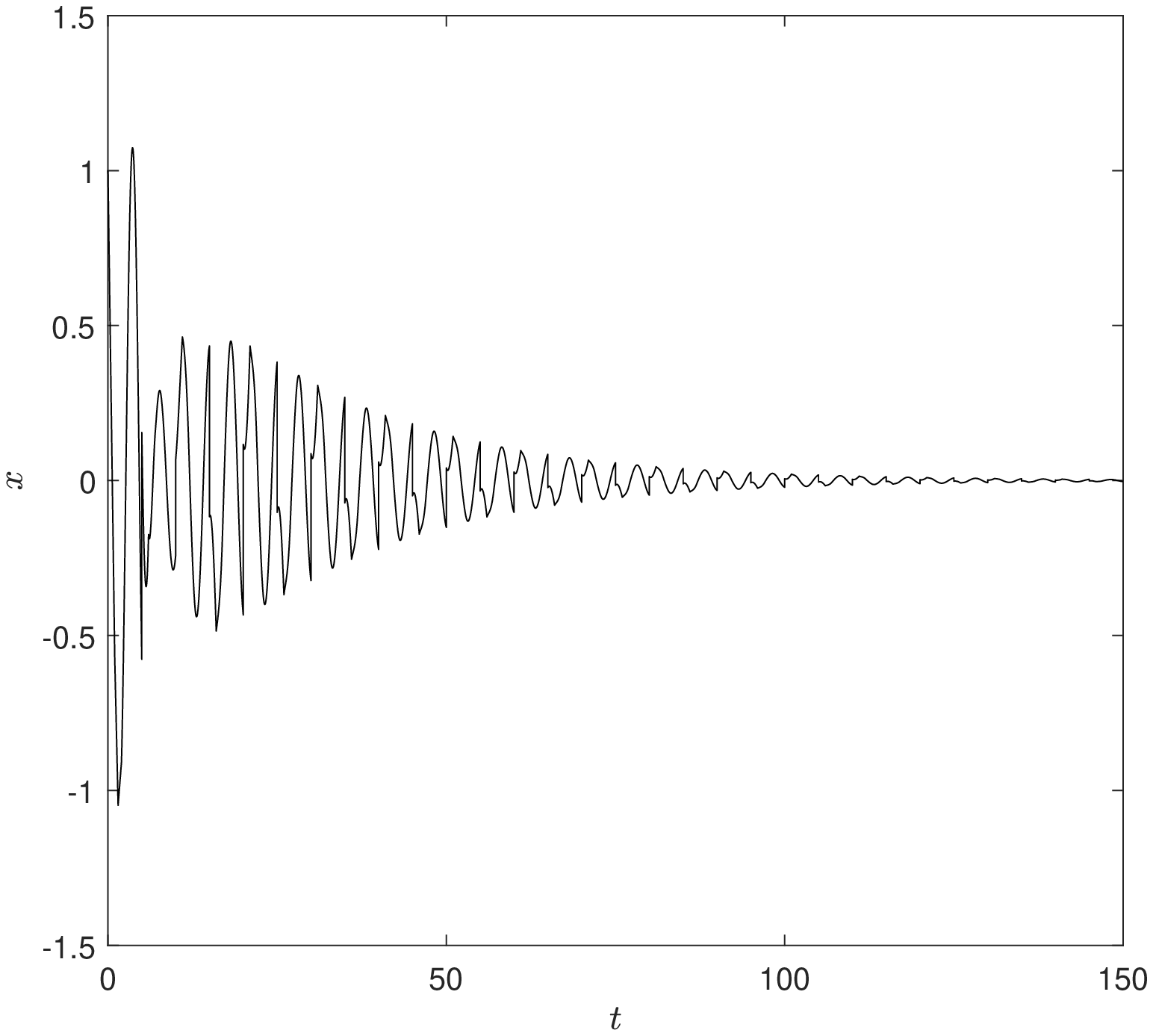}
\caption{Left: Solution of \eqref{ex3-dde1}--\eqref{ex3-dde2} with $b=-0.4837$ and $h=1$ from the constant initial condition $x_0=1$. Right: with $b=-1.2685$ and $h=1/5$. In both cases asymptotic stability is achieved.}\label{fig-ex3-1}
\end{figure}

\begin{figure}
\centering
\includegraphics[scale=0.45]{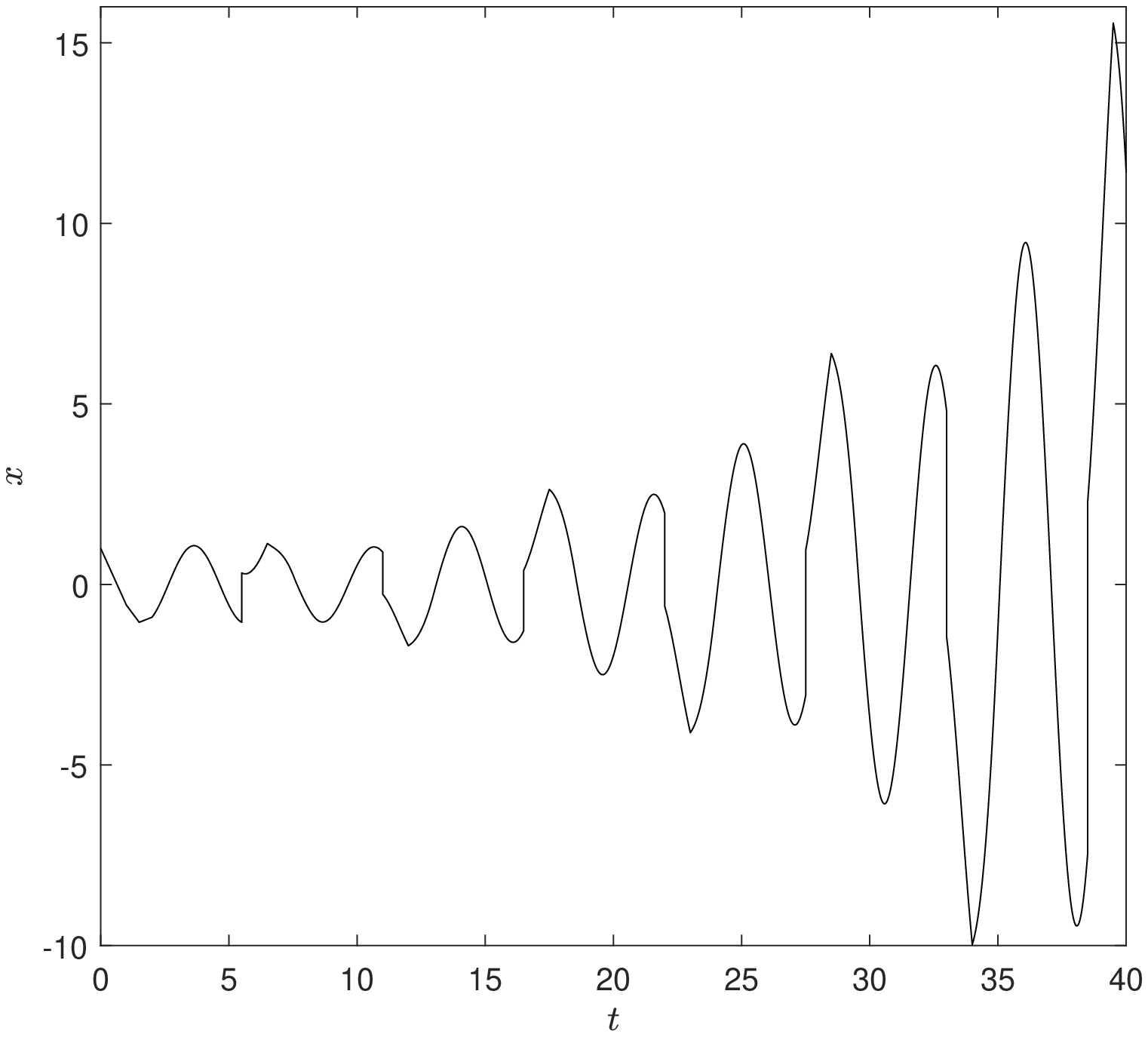}\includegraphics[scale=0.45]{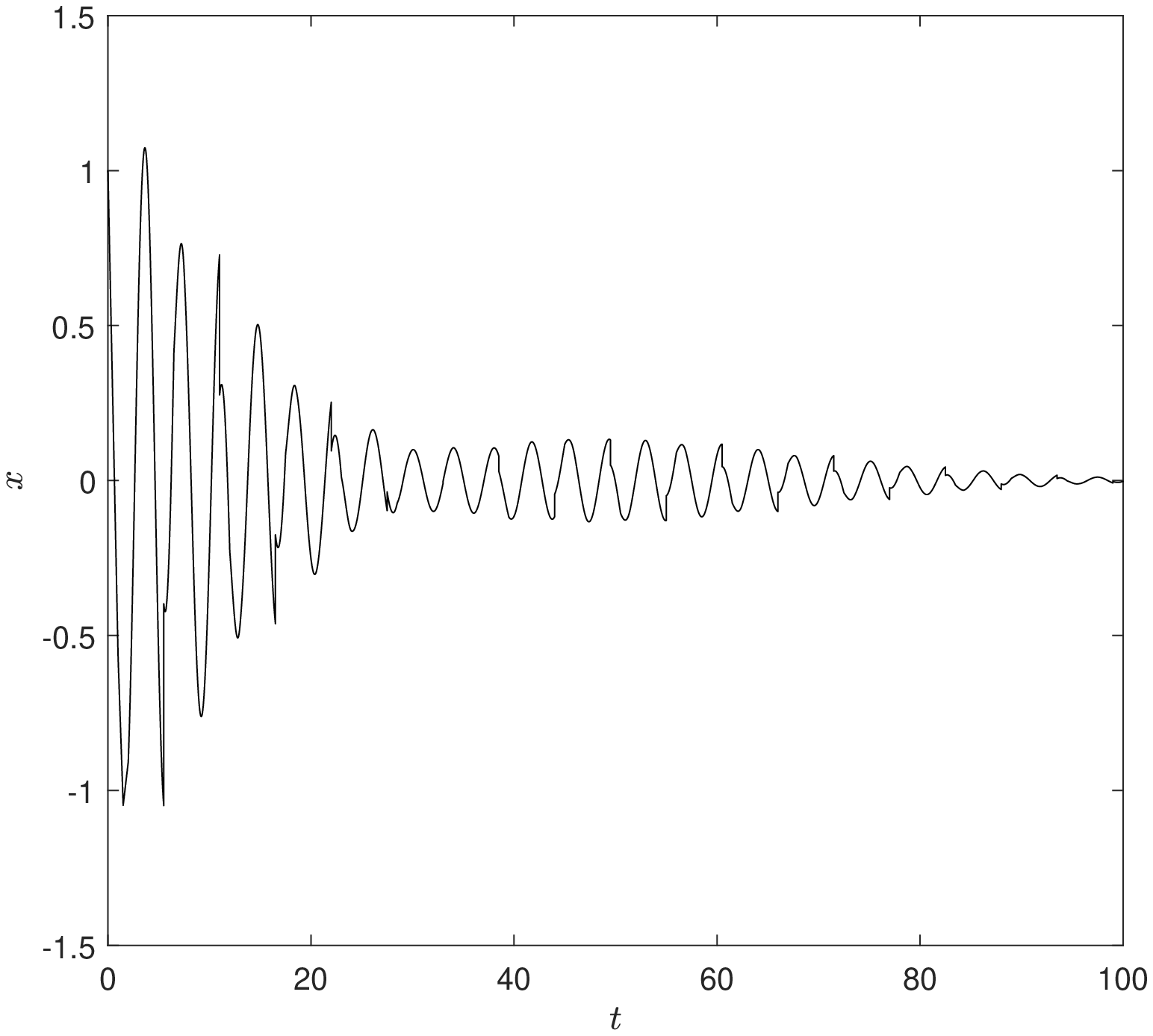}
\caption{Left: Solution of \eqref{ex3-dde1}--\eqref{ex3-dde2} with $b=-1.3045$ and $h=2/11$ from the constant initial condition $x_0=1$. Right: with $b=-0.6207$ and $h=2/11$. }\label{fig-ex3-2}
\end{figure}

We therefore can not expect the script \texttt{ISIM1s\_run} to generate a stabilizing controller if we choose $h=\frac{1}{2N}$ for some $N\in\N$. Let us perform a test. With the user input\\
\\
\noindent \texttt{A0 = 0;\\
A1 = -pi/2;\\
tau = 1;\\
control\_basis = 1;\\
h = 1/2;\\
gamma = -0.2;\\
constraint = [];\\
cost\_weight = 1;\\
cost\_general = [];\\
searchmode = `fmincon';\\
guess = 0;\\
options = [];\\
eig\_lower = -1;\\
eig\_upper = inf;\\
spectrum\_discretization = 10;\\
refining = 'false';\\}
\\
We then run the script and obtain the somewhat inconclusive output\\
\\
\noindent \texttt{>> ISIM1s\_run\\
Found 2 eigenvalues with real part at least -1.000000 for the input DDE, counting multiplicity.\\
\\
M0 of deficient rank; unconstrained problem might be infeasible with specified control space.\\
\\
Dimension of control space (1) is at most probe space dimension (4).\\
\\
Performing optimization in control space.\\
\\
Converged to an infeasible point.\\
\\
>> control\{2\}\\
ans = -8.8584e-09\\}
\\
Since we know that no controller can stabilize this system at the frequency $h=\frac{1}{2}$, it makes sense that the optimization solver did not converge. The constraint function has no feasible solution. Let us instead use the frequency parameter $h=1$. In this case the output is a bit more encouraging.\\
\\
\texttt{>> ISIM1s\_run\\
Found 2 eigenvalues with real part at least -1.000000 for the input DDE, counting multiplicity.\\
\\
M0 of deficient rank; unconstrained problem might be infeasible with specified control space.\\
\\
Dimension of control space (1) is at most probe space dimension (4).\\
\\
Performing optimization in control space.\\
\\
Local minimum found that satisfies the constraints.\\
\\
>> control\{2\}\\
ans = -0.4837\\}
\\
See Figure \ref{fig-ex3-1}; this controller does indeed ensure asymptotic stability.
What if we instead use the frequency $\texttt{h}=\texttt{1/5}$? The feasibility is not ruled out by our previous observation because the denominator $3$ is not even. In this case we get the output\\
\\
\noindent\texttt{>> control\{2\}\\
ans = -1.2685\\}
\\
Yet again, we get local asymptotic stability (although with a seemingly lower convergence rate) as required; see Figure \ref{fig-ex3-1}. What about if we further decrease the frequency to $\texttt{h}=\texttt{2/11}$? Making this adjustment to the script and running \texttt{ISIM1s\_run}, one gets the control output\\
\\
\noindent \texttt{>> control\{2\}\\
ans = -1.3045\\}
\\
The resulting simulation shows instability; see Figure \ref{fig-ex3-2}. Referring back to Section \ref{section-fail}, \texttt{ISIM1s} generally fails if the controller that is found is too large. It is possible that for the problem at hand, the specified convergence rate parameter $\texttt{gamma}=\texttt{-0.2}$ results in the feasible region being far away from zero. To remedy this, let us relax our requirement on the convergence rate and set $\texttt{gamma}=\texttt{-0.05}$.  Running \texttt{ISIM1s\_run} with this new parameter, we get the control output\\
\\
\noindent \texttt{>> control\{2\}\\
ans = -0.6207\\}
\\
This time, we achieve stability; see Figure \ref{fig-ex3-2}. The main point to take away from this example is that for some stabilization problems, it might not be possible to stabilize the system at a prescribed frequency. Changing the frequency parameter \texttt{h} -- in particular, making it larger -- can improve the result. In tandem with this parameter, the convergence rate parameter \texttt{gamma} influences the size of the feasible region. If it is too negative, the set of feasible controllers might be too large for the heuristic of \texttt{ISIM1s} to function correctly. Setting \texttt{gamma} closer to zero (but still negative) can improve feasibility.

\subsection{Work reassignment in a theoretical computing system with errors}
\emph{Note: Some of the inputs involved in this example are generated randomly, namely the matrices $R$, $E$ and vector $z$. For reproducibility, I have included the MATLAB workspace file containing those that were used to produce the simulations. Also, building the control basis for this example is a bit more involved and some of the user inputs are nonstandard. To further facilitate reproducibility, see the folder \emph{\texttt{example\_computing}} within the package zip file. This contains the workspace, the \emph{\texttt{ISIM1s\_run}} script with correct user inputs as well as a modified version of the simulation function that takes the additional input \emph{\texttt{z}}.}

Suppose $k=1,\dots,N$ worker units independently process $N$ input signals arriving at time-varying (but bounded) bit rates $s_k:\R\rightarrow\R^+$. Assuming processing occurs in $O(\log M)$ time where $M$ is the task size, each worker has an intrinsic per-bit processing rate $r_k>0$ in units of 1/time. If each worker has probability $e_k\in[0,1]$ of producing an error at each computation step, errors are identified in $\tau$ time units and on error identification the incorrect bits are returned to
the worker for processing, the rate of change in data in worker $k$'s data buffer is
\begin{align}\label{ex4-eq1}\dot x_k(t) = s_k(t) + r_k(-x_k(t)+e_kx_k(t-\tau)).\end{align}
\textbf{Disclaimer:} The author recognizes that this derivation is fraught with technical problems and this is a completely unrealistic model of computing. That said, it will be an interesting model to study for the purposes of \texttt{ISIM1s} and will provide an illustrative example on how to input a slightly nonstandard control basis.

The \emph{performance} of this computing system can be characterized in terms of the eigenvalues of the linear system
\begin{align*}
\dot x_k&=r_k(-x_k(t)+x_k(t-\tau)),&k=1,\dots,N.
\end{align*}
Namely, if $\lambda$ is the eigenvalue of this system with maximum real part and $\Re(\lambda)<0$, then the worst case amount of time needed to process a (vector) task of size $M$ (in the supremum norm) is $|\Re(\lambda)|^{-1}\log(M)$ as $M\rightarrow\infty$. For this reason, we define the performance $\kappa$ to be precisely 
$$\kappa = \left\{\begin{array}{cc}|\Re(\lambda)|,&\Re(\lambda)<0 \\ 0 & \Re(\lambda)\geq 0. \end{array}\right.$$  
When all workers are error-free -- that is, $e_k\equiv 0$ -- the performance is precisely $\kappa = \min_k\{r_k\}$. When there are errors, this performance will certainly decrease. As for the original system \eqref{ex4-eq1}, the performance influences how quickly each worker's data buffer usage will converge to a particular bounded trajectory (determined by the input signals $s_k$). The worst case convergence rate is $O(e^{-\kappa t})$ provided the performance is positive.

We are interested in whether or not it is possible to improve the performance in the case where there are error-prone workers -- that is, when $e_k>0$ for some workers -- by reassigning work. To be precise, first define $R=\mbox{diag}(r_1,\dots,r_N)$ and $E=\mbox{diag}(e_1,\dots,e_N)$. We consider the impulsive system
\begin{align}
\label{ex4-eq2}\dot x&=R\left(-x(t)+Ex(t-\tau)\right),&t&\notin\frac{1}{h}\Z\\
\label{ex4-eq3}\Delta x&=B x(t^-),&t&\in\frac{1}{h}\Z,
\end{align}
where $B$ is a $N\times N$ matrix whose diagonal entries $B_{ii}$ satisfy $B_{ii}\in[-1,0]$, all off-diagonal entries are nonnegative, and such that each column sums to zero. In effect, $B$ describes the reallocation of data among the workers. The column sum condition ensures that no data is lost/deleted, while the diagonal/off-diagonal condition ensures that the direction of flow of data is correct and no worker can transfer more data than it has available. $h$ is, as usual, the frequency of impulse effect. We ask whether we can find some $B$ satisfying these constraints such that the convergence rate is $O(e^{-\gamma t})$ for some $\gamma<\kappa$. This would amount to an improvement in performance.

First, we generate some theoretical processing rates and error probabilities. Our system will consist of $N=30$ workers, the processing rates will be drawn from the standard folded normal distribution, the error probabilities will be drawn from the uniform distribution on $[9\cdot 10^{-5},1.1\cdot 10^{-4}]$.\\
\\
\noindent \texttt{>> R = diag(abs(randn(30,1)));\\
>> E = diag(1E-4 + 1E-5*rand(30,1));\\
>> max(riag(R))\\
ans = 1.8140\\
>> mean(diag(R))\\
ans = 0.6026\\}
\\
We will select a performance target of $\kappa^* = 0.6026$, which corresponds to $\gamma=-0.6026$. This was derived from the mean performance assuming no errors. As for the eigenvalue control interval, we will use \texttt{[eig\_lower,eig\_upper] = [-1.8140,0]}, the rationale being that assuming no errors, the lower bound is the negative of the maximum performance, while the upper bound corresponds to zero performance. Under the assumptions on the matrices $R$ and $E$, individual nodes are very likely to have positive performance.

Introduce the matrix $B(i,j)$ for $i\neq j$, whose $j$th diagonal entry is $-1$, the $(i,j)$ entry is $+1$ and all other entries are zero. The set
\begin{align*}
\mathcal{U} = \bigcup_{j=1}^N\{B(i,j) : i\in\{1,\dots,N\}\setminus\{j\}\}.
\end{align*}
consists of matrices with only two nonzero entries, such that one of the diagonal elements is $-1$ and some entry above or below is $+1$. It therefore provides a basis for the set of matrices with zero column sum. With $N=30$ the dimension of $\mbox{span}(\mathcal{U})$ is $D=30\cdot29=870$. 

To accommodate for the restriction that the main diagonal entries of our control matrices $B$ must be in the interval $[-1,0]$ and any other nonzero entries must be positive, we will specify an ordering on the basis elements $B_k$ such that $B_1,\dots,B_{29}$ have $B[1,1]=-1$, the elements $B_{1+29},\dots,B_{2\cdot29}$ have $B[2,2]=-1$ and so forth, where square braces denote row-column indices. Then, in terms of the associated coordinate vector $x\in\R^{D}$, the control space is specified by the linear constraints
\begin{align*}
x&\geq 0\\
\mbox{diag}(\mathds{1},\dots,\mathds{1})x&\leq 1,
\end{align*}
where $\mathds{1}$ is the $1\times 29$ row vector of ones and the block diagonal matrix is formed of $1\times 29$ blocks, so that the matrix above has dimensions $30\times D$. The inequalities are considered componentwise.

\begin{figure}
\centering
\includegraphics[scale=0.7,trim={3cm 0 3cm 0}]{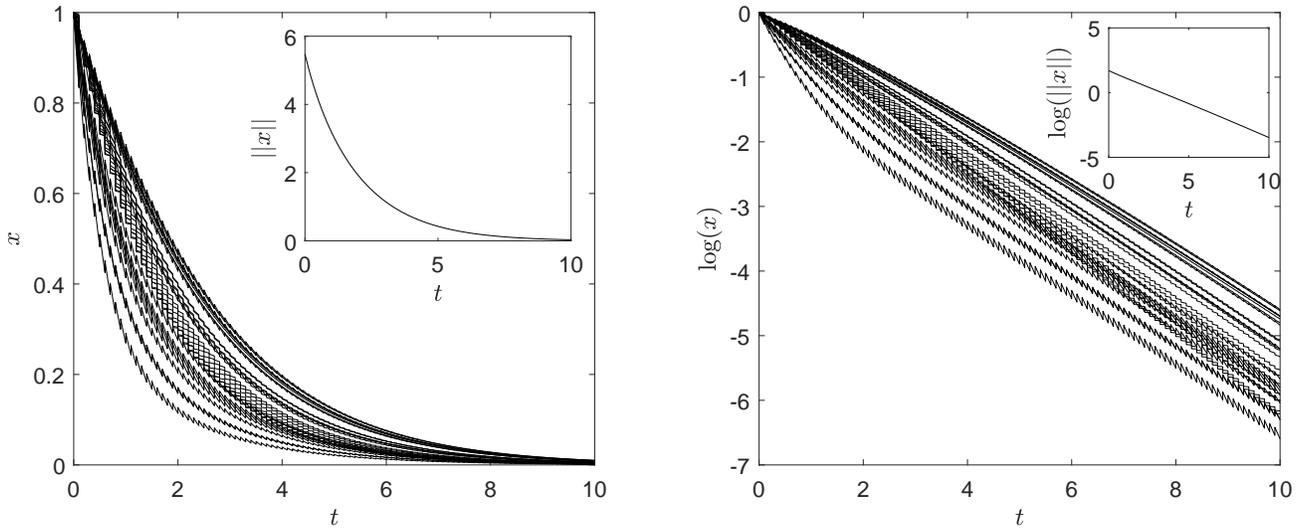}
\caption{Solution of \eqref{ex4-eq2}--\eqref{ex4-eq3} with the controller $B=\texttt{control\{2\}}$ from the constant initial condition $x_0=\texttt{ones(30,1)}$. Left: all components plotted; inset with solution norm plotted. Right: all components in logarithmic scale; inset with logarithm of solution norm. Observe the (approximate) linearity in the latter inset log-norm plot.}\label{fig-ex4-1}
\end{figure}

We will take the control frequency to be $h=10$ and the error detection delay to be $\tau=0.2$. To keep things simple we will yet again take the standard unweighted quadratic cost functional. It will turn out that the convergence the optimization step is slow and, with the dimension being high, requires many function evaluations. We will use the options structure to allow up to $10^5$ function evaluations\footnote{Even this will not be enough for the solver to converge, but as we will see the output is at least feasible.} using the (fastest) \texttt{`fmincon'} smooth solver. With the stock spectrum discretization level, our user input to \texttt{ISIM1s\_run} is\\
\\
\noindent\texttt{
A0 = -R;\\
A1 = R*E;\\
tau = 0.2;\\
\% -- build the control basis -- \\
control\_array = cell(1,30*29);\\
for i=1:30\\
    \indent for j=1:29\\
        \indent\indent if i+j<=30\\
            \indent\indent\indent control\_array\{(i-1)*29 + j\}=zeros(30,30);\\
            \indent\indent\indent control\_array\{(i-1)*29 + j\}(i,i)=-1;\\
            \indent\indent\indent control\_array\{(i-1)*29 + j\}(i+j,i)=1;\\
        \indent\indent else\\
            \indent\indent\indent control\_array\{(i-1)*29 + j\}=zeros(30,30);\\
            \indent\indent\indent control\_array\{(i-1)*29 + j\}(i,i)=-1;\\
            \indent\indent\indent control\_array\{(i-1)*29 + j\}(mod(i+j,30),i)=1;\\
        \indent\indent end\\
   \indent end\\
end\\
control\_basis = cell2mat(control\_array);\\
clear control\_array\\
\% -- control basis built --\\
h = 10;\\
gamma = -0.6026;\\
\% -- build the constraint matrix -- \\
C = sparse(kron(eye(30,30),ones(1,29)));\\
\% -- constraint matrix built --\\
constraint = @(x)[-x ; C*x - ones(30,1)];\\
cost\_weight = 1;\\
cost\_general = [];\\
searchmode = `fmincon';\\
guess = zeros(30*29,1);\\
options = optimoptions(`fmincon',`MaxFunctionEvaluations',1E5);\\
eig\_lower = -1.8140;\\
eig\_upper = 0;\\
spectrum\_discretization = 10;\\
refining = `false';\\}
\\
The loops in our code are far from the most efficient way to build the control basis, but we will be content with this for now. The output from \texttt{ISIM1s\_run} is as follows:\\
\\
\noindent\texttt{>> ISIM1s\_run\\
Found 30 eigenvalues with real part in the interval [-1.814000,0.000000] for the input DDE, counting multiplicity.\\
\\
Dimension of control space (870) is at most probe space dimension (900).\\
\\
Performing optimization in control space.\\
\\
Solver stopped prematurely.\\}
\\
The solver stopped prematurely because it completed $10^5$ function evaluations, which was our specified maximum. This run took 149.607 seconds on a Ryzen 5 1500X. To compare the performance of \eqref{ex4-eq2}--\eqref{ex4-eq3} with and without the impulsive controller -- that is, with $B=\texttt{control{2}}$ and with $B=0$ --  we first need the performance of the system without impulses, \eqref{ex4-eq1}. This can be found using the last entry of \texttt{eigs\_all}. \\\
\\
\noindent \texttt{>> flip(eigs\_all(end-6:end))\\
ans = [-0.0458;-0.0638;-0.1093;-0.1096;-0.1178;-0.1269;-0.1555]\\}
\\
The (worst-case / asymptotic) performance without the controller is therefore $\kappa_0=0.0458$. To determine the actual asymptotic performance of the system with the controller $B=\texttt{control\{2\}}$, we will use linear regression on the logarithm of the norm of the solution generated by \eqref{ex4-eq2}--\eqref{ex4-eq3}. See Figure \ref{fig-ex4-1}.
Basic fitting of the log-norm plot produces a linear fit with slope $-0.60483$ with norm of residuals $0.55837$. An estimate for the performance is therefore $\tilde\kappa = 0.60483$, which is actually better than our target of $\kappa^*=0.6026$. This might be because the solver stopped prematurely; convergence to the boundary $\rho(\mathcal{M})=e^{\gamma/h}$ did not occur and so the cost was not minimized.

\begin{figure}
\centering
\includegraphics[scale=0.7,trim={3cm 0 3cm 0}]{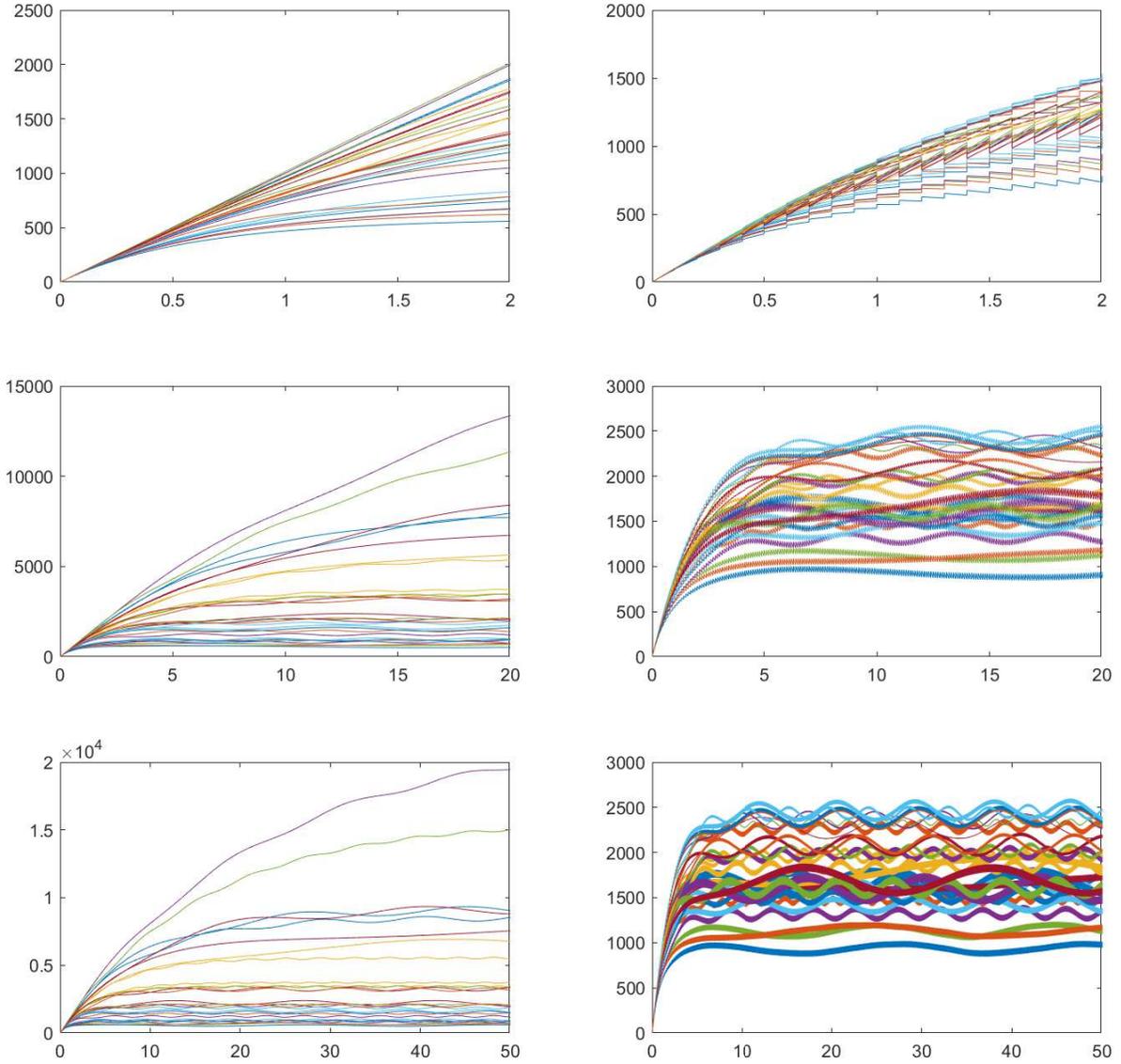}
\caption{Solution of \eqref{ex4-eq1} with the controller determined by the jump condition \eqref{ex4-eq3}, from the constant initial condition $x_0=\texttt{zeros(30,1)}$ and with the inputs $s_k(t)=1000\cdot(1+0.1\cdot\sin(tz_k))$ with $z_k\sim N(0,2\pi)$. Left: no controller. Right: with the controller $B=\texttt{control\{2\}}$ generated by \texttt{ISIM1s}. Time is on the horizontal axis, with worker state (data buffer usage) in the vertical axis. Observe that aside from yielding faster convergence to the quasiperiodic steady state, the controller has the effect of more evenly distributing work relative to the workers' processing speeds and error rates, as indicated by the smaller variance of data buffer usage across workers.}\label{fig-ex4-2}
\end{figure}

To demonstrate the performance of the controller with some nontrivial inputs, we will simulate \eqref{ex4-eq1} with input signals of the form $s_k(t)=1000\cdot(1+0.1\cdot\sin(z_kt))$ for $z_k$ a sequence of normal random variables with mean zero and variance $2\pi$ (this can be generated with \texttt{z = sqrt(2pi)*randn(30,1)}) and all data buffers initially empty (that is, $x_0=0$). We will simulate the system both with $B=0$ and with $B=\texttt{control\{2\}}$. This is provided in Figure \ref{fig-ex4-2}. At $50$ time units the system without control still has some data buffers that have not reached their steady state oscillatory response and are still growing, whereas in the system with control the workers have settled into a stable quasiperiodic oscillation.

\subsection{Stabilization of a high-dimensional nonlinear network}\label{ex5}
\emph{Note: Like the previous example, this one contains data that is randomly generated. See the folder \emph{\texttt{example\_network}} within the zip file of this package for a MATLAB workspace and \emph{\texttt{ISIM1s\_run}} script containing relevant user inputs.}

We will consider the nonlinear network model/control problem
\begin{align}
\label{ex5-eq1}\dot x_i&=-x_i(t) + \left[V\tanh(x_i(t)) + W\tanh(x_i(t-1))\right] + \sum_{j=1}^N a_{ij}x_j(t),&t&\notin\Z\\
\label{ex5-eq11}\Delta x_i&=B_ix_i(t^-),&t&\in\Z,
\end{align}
for two-dimensional nodes $x_i$ with $i=1,\dots,N$, $\tanh$ the componentwise hyperbolic tangent function, matrices $V$ and $W$ affecting the interval dynamics given by
$$V=\left[\begin{array}{cc}2 & -0.11\\-5 & 3.2 \end{array}\right],\hspace{1cm}W=\left[\begin{array}{cc}-0.8&-0.05\\-0.09&-1.2\end{array}\right],$$ and $A=(a_{ij})_{N\times N}$ the negative of a graph Laplacian associated to a small world network graph on $N$ nodes. Each of the $B_i$ are $2\times 2$ matrices representing the impulsive control. When there is no coupling ($A=0$), the individual nodes of this delay differential equation each have $x-0\in\R^2$ as an unstable equilibrium. In this example we will use $N=100$ nodes.

Performing a linearization of \eqref{ex5-eq1} at the origin $x_i\equiv 0$, one can write the resulting system in matrix form
\begin{align*}
\dot y&=(I_{N\times N}\otimes(-I_{2\times 2}+V)+A\otimes I_{2\times 2})x(t) + (I_{N\times N}\otimes W)x(t-1),
\end{align*}
where $\otimes$ denotes the Kronecker product. We will attempt to stabilize $y=0$ by designing $B=\mbox{diag}(B_1,\dots,B_{100})$ with $B_i\in\R^{2\times 2}$ such that 
\begin{align}
\label{ex5-eq2}\dot y&=(I_{N\times N}\otimes(-I_{2\times 2}+V)+A\otimes I_{2\times 2})x(t) + (I_{N\times N}\otimes W)x(t-1),,&t&\notin\Z\\
\label{ex5-eq3}\Delta z&=Bz(t^-),&t&\in\Z,
\end{align}
is asymptotically stable. Since this particular system has both a large amount of coupling and is rather high-dimensional, we will give \texttt{ISIM1s} a bit of help\footnote{Our test system will have a large positive real eigenvalue 2.4704. Stabilizing such a system really is not in the scope of \texttt{ISIM1s}, since this large real eigenvalue is in all likelihood too far away from the imaginary axis for a small controller to stabilize it (or a large controller is quite likely to introduce additional instability). Hence, we will structure our control space to give our heuristic the greatest chance at success.} by only searching for explicitly diagonal controllers. That is, we will take the control space to be (a subset of) $$\mathcal{U}=\{B\in\R^{2N\times 2N} : B[i,j]=0, i\neq j\},$$ the diagonal matrices. We will also only allow controllers that induce negative feedback, which is to say that we will impose the additional constraint that these matrices have diagonal entries in the interval $[-1,0]$. With $N=100$ nodes, the control space can be identified with a convex subset of $\R^{200}$, and each of these $e_k\in\R^{200}$ standard ordered basis vectors will be identified with the (diagonal) basis element having a one in its $(k,k)$ entry, and zeros everywhere else.

To make things a bit more interesting, we will use the cost function
$$C(x)=\frac{1}{\max\{D[i,i]:i=1,\dots,N\}}x^\intercal (D\otimes I_{2\times 2})x + \sum_{k=1}^{200}\tanh(4|x_k|).$$
Here, $D\in\R^{100\times 100}$ is the diagonal matrix consisting of the in-degrees of the nodes of $G$. The first term causes nodes with higher degree to be prescribed a higher cost, weighted by the maximum degree. The second term is essentially an approximation to a discontinuous penalty for each nonzero control element (relative to the chosen basis).

\begin{figure}
\centering
\includegraphics[scale=0.6,trim={4cm 0 4cm 0}]{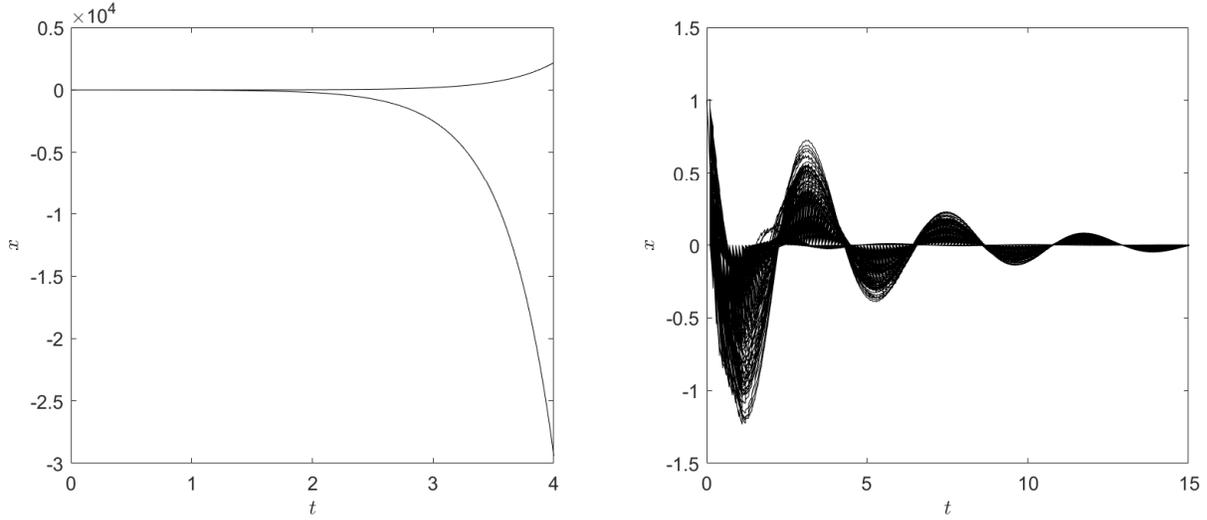}
\caption{Plots of the solution of \eqref{ex5-eq2}--\eqref{ex5-eq3} through the constant initial condition $x_0=\texttt{ones(200,1)}$. All 200 individual solution components are plotted. Left: $B=0$. Right: with $B=\texttt{control\{2\}}$.}\label{fig-ex5-1}
\end{figure}

\begin{figure}
\centering
\includegraphics[scale=0.6,trim={4cm 0 4cm 0}]{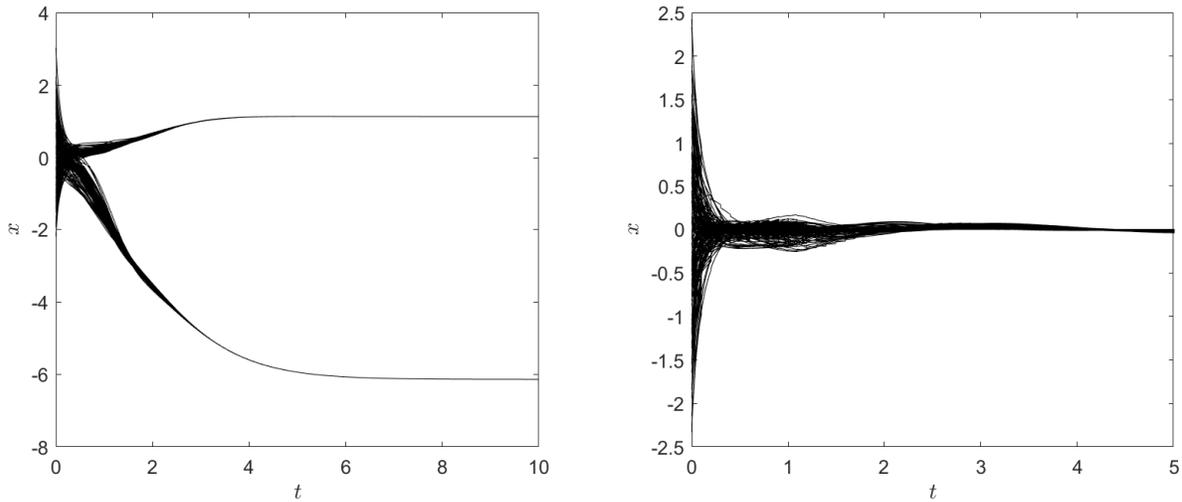}
\caption{Plots of the solution of \eqref{ex5-eq1}--\eqref{ex5-eq11} through standard normally distributed constant initial condition $x_0=\texttt{randn(200,1)}$. All 200 individual solution components are plotted. Left: $B=0$. Right: with $B=\texttt{control\{2\}}$. Notice that when $B=0$ (left) the solution converges to a nonzero steady state.}\label{fig-ex5-2}
\end{figure}

Running the included \texttt{ISIM1s\_run\_networkexample} script, the result is a diagonal matrix \texttt{control\{2\}} that guarantees exponential stability of the linearization \eqref{ex5-eq2}--\eqref{ex5-eq3} and, subsequently, local asymptotic stability of the full nonlinear system \eqref{ex5-eq1}--\eqref{ex5-eq11}. See Figure \ref{fig-ex5-1} and \ref{fig-ex5-2} for plots of both the linear and nonlinear dynamics from the usual initial condition, with and without the control.

We should remark that the included demo script \texttt{ISIM1s\_run\_networkexample} uses the convergence rate parameter $\gamma=0$, yet to linear order the dynamics with the resulting controller feature exponential decay (i.e.\ the convergence rate parameter achieved is actually negative). This might seem contradictory, but it is really more of an indication that the heuristic of \texttt{ISIM1s} does not work correctly for this example system. The fact that the algorithm uses only a \emph{single stage} of linearization is a serious problem.

\subsection{General conditioning guidelines and error messages}\label{conditioning}
To keep this section as simple as possible, what follows are some general guidelines you might follow if \texttt{ISIM1s} fails to stabilize your system. There are also a few general comments concerning possible explanation for certain errors that might come up.\\
\\
\emph{Conditioning.}
\begin{itemize}
\item Decreasing \texttt{eig\_lower} to cover some (additional) negative eigenvalues can resolve issues where an eigenvalue with negative real part might be destabilized by the controller produced by \texttt{ISIM1s}. View all your eigenvalues with \texttt{eigs\_all} to get an estimate for how many have positive real part and how many have negative real part but are close to the imaginary axis.
\item It is advised to keep $\gamma\leq 0$ small, at least initially. Taking $\gamma$ too large can result in a large controller for which the heuristic of \texttt{ISIM1s} might not work.
\item Conversely, if your target convergence rate is not reached (e.g.\ you wanted $\gamma=-0.8$ but only observed\footnote{Observed convergence rate parameters can be estimated by linearly fitting a log-norm plot, such as in Figure \ref{fig-ex4-1}} $\gamma=-0.3$) but stabilization has been achieved, consider making $\gamma$ more negative.
\item Increasing the frequency $h$ can improve feasibility because (generally) a smaller controller will be found.
\item If your problem is very high-dimensional and involves lots of linear coupling, it can help to impose negative feedback conditions on your control space. Section \ref{ex5} provides a good example of this.
\item If \texttt{ISIM1s} fails, try changing the frequency parameter slightly -- even by a few decimal places. Although unlikely, your problem might be one of those (see Section \ref{ex3}) at which for some frequencies, impulsive stabilization is not possible (given your control space).
\end{itemize} 

\noindent\emph{Error messages.}
\begin{itemize}
\item If \texttt{dde\_data} complains about a badly scaled matrix, try increasing the \texttt{spctrum\_discretization} parameter. This error may be due to imprecise numerical eigenvalues/eigenvectors from the DDE.
\item If either of \texttt{fmincon} or \texttt{patternsearch} solvers converge to an infeasible point, try a different guess. For convenience, \texttt{guess = []} will use a random input. If using \texttt{ga}, run the script again. In these cases, set \texttt{refining = `true'} so that \texttt{dde\_data} is not run again.
\item If the solver says that no feasible point was found, try:
\begin{itemize}
\item increasing the frequency $h$;
\item putting $\gamma$ closer to zero;
\item increasing the dimension of your control space or relaxing some constraints;
\end{itemize}
\item Any issues with dimension mismatch from \texttt{fmincon} or one of the other optimization solvers is likely due incompatibility between the dimensions of \texttt{control\_basis}, your DDE data \texttt{A0} (which is used to infer the dimension $n$ of your problem), any of the cost functions/weights and constraint functions. Check these thoroughly to ensure that they are all consistent.
\end{itemize}
\bibliographystyle{plain}
\bibliography{C:/Users/churc/Sync/AcademicDocuments/Mendeley/BibTex/library}

\begin{thebibliography}{1}

\bibitem{Church2019a}
Kevin Church.
\newblock {\em {Invariant manifold theory for impulsive functional differential
  equations with applications}}.
\newblock PhD thesis, University of Waterloo, 2019.

\bibitem{Church_IEETAC}
Kevin Church and Xinzhi Liu.
\newblock {Invariant manifold-guided impulsive stabilization of delay
  equations}.
\newblock {\em IEEE Transactions on Automatic Control}, pages 1--1, 2021.

\bibitem{Church2019b}
Kevin E~M Church and Xinzhi Liu.
\newblock {Cost-Effective Robust Stabilization and Bifurcation Suppression}.
\newblock {\em SIAM Journal on Control and Optimization}, 57(3):2240--2268, jan
  2019.

\bibitem{ChurchLiu_monograph}
Kevin E.~M. Church and Xinzhi Liu.
\newblock {\em {Bifurcation Theory of Impulsive Dynamical Systems}}.
\newblock Springer International Publishing, 2021.

\bibitem{Jarlebring2008}
Elias Jarlebring.
\newblock {Some numerical methods to compute the eigenvalues of a time-delay
  system using MATLAB}.
\newblock {\em The delay e-letter}, 2, 2008.

\bibitem{Trefethen2000}
Lloyd~N. Trefethen.
\newblock {\em {Spectral Methods in MATLAB}}.
\newblock Society for Industrial and Applied Mathematics, jan 2000.

\end{thebibliography}

\end{document}